\documentclass{article}
\usepackage{pgfplots}
\usepackage{lipsum}
\usepackage{caption}
\usepackage{subcaption}
\usepackage{authblk}
\usepackage[english]{babel}
\usepackage{amsfonts}
\usepackage{amsmath}
\usepackage[top=2cm, bottom=2cm, left=2cm, right=2cm]{geometry}
\usepackage{fancyhdr}
\usepackage{multicol}
\usepackage{bbold}
\usepackage{epsfig}
\usepackage{graphics}
\usepackage{epsfig}
\usepackage{lipsum}
\usepackage{fancyhdr}
\usepackage{booktabs}
\usepackage{cleveref}
\usepackage{wrapfig}
\usepackage[export]{adjustbox}
\usepackage{rotating}
\usepackage{amsmath}

\renewenvironment{abstract}{
	
	\hfill\begin{minipage}{0.95\textwidth}
		\rule{\textwidth}{1pt}}
	{\par\noindent\rule{\textwidth}{1pt}\end{minipage}
}


\begin{document}

	\title{\textbf{Strong quantum correlation in a pair hybrid optomechanical cavities }}
	\author[1]{\textbf{Khadija El Anouz}}
	\author[1,2]{\textbf{Abderrahim El Allati}}
	\author[*,4, $\dagger$]{\textbf{Farhan Saif}}
	\affil[1]{\small Laboratory of R\&D in Engineering Sciences, Faculty of Sciences and Techniques Al-Hoceima, Abdelmalek Essaadi University, Tetouan,
		Morocco}
	\affil[2]{\small  The Abdus Salam International Center for Theoretical Physics,
		Strada Costiera 11, Miramare-Trieste, Italy}
	\affil[*]{\small	Department of Electronics, Quaid-i-Azam University, 3rd Avenue, Islamabad 45320, Pakistan}
	\affil[4]{\small Department of Engineering Sciences, University of Electro-Communications, Chofu, Tokyo, Japan}
	\affil[$\dagger$]{\small fsaif@yahoo.com }
	\maketitle
	\begin{center}
		\textbf{Abstract}
	\end{center}
	\begin{abstract}
We show  the quantum correlation between two coupled  hybrid optomechanical cavities  by quantifying the non-classical correlation using Gaussian quantum discord. This involves analyzing and solving Heisenberg Langevin equations to obtain the $(12\times12)$-dimensional covariance matrix of this system. Based on the experimentalist conditions, we simulate quantum correlation of bipartite steady-state with continuous conditions using Guassian quantum discord. We know that the generation of quantum correlation and its robustness essentially depend on the physical parameters of the system. We provide the stability analysis by means of the Ruths–Hurwitz criterion to confirm the choices made during the analysis of quantum discord dynamics.

	\end{abstract}
	
	\textbf{Keywords}: Optomechanical cavities; Guassian quantum discord, Stability.
	
	\section{Introduction}
	
	Over few decades, quantum correlations  in composite quantum  systems leads to develop various fundamental  concepts of quantum information and advanced quantum communication techniques \cite{e1,e2,e3}. Quantum discord is one of the quantum correlations measures introduced by Ollivier and Zurek  \cite{e4}. There are many ways to study geometric discord which are Hellinger distance  \cite{s1}, Bures distance \cite{s2}, trace distance \cite{s3}, Hilbert-Schmidt distance  \cite{e5,e6}, Gaussian quantum discord \cite{e9}, and recently local quantum uncertainty  \cite{e7} and  local quantum Fisher information \cite{e8}. These of measures have been shown to have a significant impact on the quantification of quantum correlations in deriving their applications as a resource in quantum information processing. Generally, when the quantum discord is expanded to a two-mode Gaussian-like state, the Gaussian quantum discord \cite{e9} is a good quantifier of non-classical correlation \cite{e10}. The bipartite Gaussian state is correlated if the Gaussian discord is larger than one, while the state is either unentangled or correlated if and only if the Gaussian quantum discord is bounded between zero and one. Interestingly enough, a specific property of quantum correlation is given in the possibility of transmitting information securely between two partners using the quantum teleportation protocol. When implementing a quantum teleportation scheme using a partially or maximized quantum channel, it is necessary to estimate the quality of this protocol by quantifying the fidelity measure, which allows us to assess the similarity between the input and output states during this process \cite{e11}. \\
	
	Recently, optomechanical cavities have attracted great attention and created an important addition in quantum technologies because of their experimental realization. These systems  interactions  of optical and mechanical resonators. Recent experiments involving micro as well as nano-mechanical resonators have proven themselves important in the nano devices \cite{e14}. Furthermore, the matter-light joint resonator in optomechanical systems is  shown to produce a high-finesse cavity due to  the interaction between an array of atoms and a mode of light. A large optomechanical coupling force, which allows the quantum dynamics of moving end-mirror to be controlled via the intracavity field \cite{e15}. Interestingly, in the case where a cavity interacts with a collective density excitation of ultracold atoms, there is a large coupling force \cite{e16}. As a result, the combination of cavity quantum electrodynamics. The ultracold gases has enabled a new applications in optomechanical systems \cite{e17, e18}.  \\
	
	The goal of this paper is to investigate non-classical  correlation by means of Gaussian quantum discord. The analysis  of the Langevin equations allows to the computation of the covariance matrix of two coupled hybrid optomechanical cavities which made up of  two identical Fabry-Pérot cavities of length $L$ with a moving end mirror induced by a single-mode optical field. Each hybrid optomechanical cavity is interacted with a Bose-Einstein Condensates (BEC), with N atoms trapped in an optical lattice potential \cite{Mikaeili}. Moreover, we suppose that the moving end mirror for each cavity reveals Brownian motion in the absence of coupling with radiation pressure. The obtained covariance matrix is a $(12\times12)$-dimensional matrix and for sake of simplicity we suppose that the cavities are identical. However, in order to investigate quantum correlation by means of Gaussian quantum discord,  we shall illustrate from this matrix seven $(4\times4)$ submatrices. Indeed, the block covariance matrices, are obtained from the interaction between first intracavity photon-phonon, second intracavity photon-phonon,  cavity modes, mechanical resonator modes, BEC-first mechanical mode system, BEC-second mechanical mode system, BEC-first optical mode system. Finally, the last block examines BEC-second  mechanical mode system.\\
	
	The paper is structured as the following:  in Sec. 2 we give a detailed description of the proposed model. Sec. 3 includes our calculations used to investigate the Langevin equations and covariance matrix related to the proposed model. In Sec. 4  we gives some preliminaries of Gaussian quantum discord and a discussion of our results. Finally, we conclude our results present  and some future directions in Sec. 5.
	
	\section{Description of the model}
	
	Let's consider a hybrid optomechanical system that allows us to generate quantum correlations in continuous variables between the intracavity optical and mechanical modes. We treat  two coupled optomechanical cavities, namely $A$ and $B$. The optomechanical cavity $A$ ($B$) is a Fabry-Pérot cavity of length $L$ with a moving end mirror represented by a single-mode optical field of frequency $\omega_{p}^{A}$ ($\omega_{p}^{B}$). Furthermore, the moving end-mirrors exhibit harmonic oscillations denoted by $\omega_m^{i}$ ($i=A, B$), which have Brownian motion in the absence of coupling with radiation pressure. Our main goal is to examine the quantum correlations generated between these cavities. To achieve this goal, we assume that the hybrid optomechanical cavites are interacted with a BEC, with N atoms trapped in an optical lattice potential \cite{Dalafi} under the condition of far of resonance. Moreover, the cavities are connected with  another via the hopping rate parameter, namely $J$ (See Fig.(1)). Finally, both fixed sides of the hybrid cavities are exposed to the output field of a squeezed light source (SLS).  \\
	
	\begin{figure}[h!]
		\begin{center}
			\includegraphics[scale=0.7]{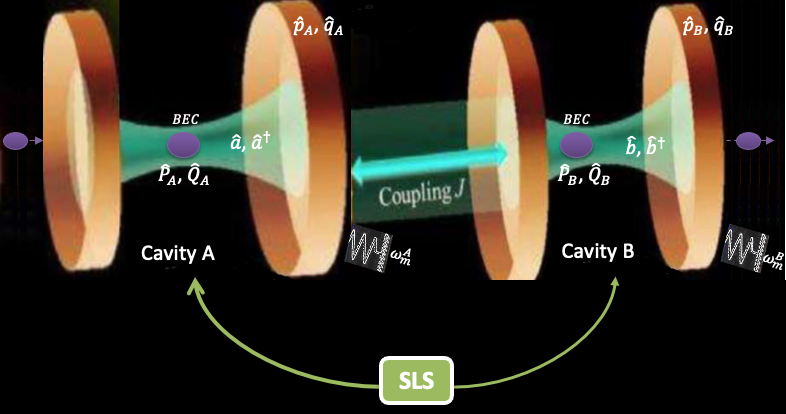}
			\caption{Schematic diagram of the joint optomechanical cavity-cavity system. The cavities are coupled with photon hopping process (PH). Moreover, the fixed sides of the hybrid cavities are exposed to the output field of a squeezed light source (SLS). Each hybrid optomechanical cavity is interacted with a BEC, with N atoms}
		\end{center}
	\end{figure}
	
	The total  Hamiltonian of the aforementioned model is given as
	\begin{equation}\label{eq1}
		\hat{H}=\hat{H}_1+\hat{H}_{2}+\hat{H}_{I}+\hat{H}_{D},
	\end{equation}
	where the Hamiltonian $\hat{H}_1$ ($\hat{H}_2$) denotes the atom-optomechanical cavity A (B). The Hamiltonian $\hat{H}_{I}$ characterizes the interaction between the cavities $A$ and $B$. While, $\hat{H}_{D}$ reports the whole effects of dissipation effects including noises and damping related to the system. These Hamiltonians are expressed as follows \cite{k1,k2, k2a}:
	\begin{eqnarray}\label{bb1}
		\hat{H}_1	&=& \hat{H}_{f_1}^{A}+\hat{H}_{m}^{A}+\hat{H}_{a}^{A}+\hat{H}_{f_2}^{A}+\hat{H}_{I_1}^{A}+\hat{H}_{I_1}^{A}, \nonumber\\
		\hat{H}_2	&=& \hat{H}_{f_1}^{B}+\hat{H}_{m}^{B}+\hat{H}_{a}^{B}+\hat{H}_{f_2}^{B}+\hat{H}_{I_1}^{B}+\hat{H}_{I_1}^{B},  \nonumber\\
		\hat{H}_I	&=& J(\hat{b}^{\dagger} \hat{a}+\hat{a}^{\dagger} \hat{b}), 
	\end{eqnarray}
	where the operators $a (a^{\dagger} )$  and $b (b^{\dagger} )$ are the annihilation (creation) oprators within cavities $A$ and $B$, respectively. Various Hamiltonians offered in  Eq.(\ref{bb1}), namely $\hat{H}_{f_1}^{A(B)}, \hat{H}_{m}^{A(B)}, \hat{H}_{a}^{A(B)}, \hat{H}_{f_2}^{A(B)}, \hat{H}_{I_1}^{A(B)}$ and $\hat{H}_{I_1}^{A(B)}$ are the Hamiltonians of the  field inside cavity $A(B)$, moving-end miror in cavity $A(B)$, atom, the field arrived from the desexcitation  of the atom inside cavity $A(B)$, the coupling between the field and the moving-end miror and the coupling between the atom and the field Hamiltonians, respectively. They take the following compact form 
	\begin{eqnarray}\label{bbb1}
		\hat{H}_{f_1}^{A}&=& \hbar  \Delta_A \hat{a}^{\dagger} \hat{a},  \nonumber \\
		\hat{H}_{f_1}^{B}&=& \hbar  \Delta_B \hat{b}^{\dagger} \hat{b},  \nonumber \\
		\hat{H}_{m}^{A(B)}&=&\frac{\hbar \omega_m^{A(B)}}{2} (\hat{p}_{A(B)}^2+\hat{q}_{A(B)}^2), \nonumber \\
		\hat{H}_{a}^{A(B)}&=& \frac{\hbar \Omega_{A(B)}}{2} (\hat{P}_{A(B)}^2+\hat{Q}_{A(B)}^2), \nonumber \\
		\hat{H}_{f_2}^{A}&=&\frac{\hbar U_0^{A} N^{A}}{2} \hat{a}^{\dagger} \hat{a},  \nonumber \\
		\hat{H}_{f_2}^{B}&=&\frac{\hbar U_0^{B} N^{B}}{2} \hat{b}^{\dagger} \hat{b},  \nonumber \\
		\hat{H}_{I_1}^{A}&=& -\zeta_A\hbar \hat{a}^{\dagger} \hat{a} \hat{q}_A-i \hbar \eta_A(\hat{a}-\hat{a}^{\dagger}), \nonumber \\
		\hat{H}_{I_1}^{B}&=& -\zeta_B\hbar \hat{b}^{\dagger} \hat{b} \hat{q}_B-i \hbar \eta_B(\hat{b}-\hat{b}^{\dagger}), \nonumber \\
		\hat{H}_{I_2}^{A}&=& \xi_m^{A} \hbar \hat{a}^{\dagger} \hat{a} \hat{Q}_A, \nonumber \\
		\hat{H}_{I_2}^{B}&=& \xi_m^{B} \hbar \hat{b}^{\dagger} \hat{b} \hat{Q}_B.
	\end{eqnarray}
	Let's note that $\Delta_{A(B)}=\omega_{A(B)}-\omega_{p}^{A(B)}$ represents the detuning parameter, such that $\omega_{A(B)}$ is the frequency associated to the cavity $A(B)$. Moreover, $\hat{p}_{i}$ and  $\hat{q}_{i}$  (i=A,B) define the dimensionless momentum and position  operators for moving end mirror. Moreover, $\omega_m^{A(B)}$ denotes the harmonic oscillations frequency of moving end mirror system. However, $\Omega_{A(B)}$ defines the recoil frequency of an atom due to the change in energy. Indeed, the atom has dimensionless momentum and position  operators, namely $\hat{P}_{A(B)}$ and $\hat{Q}_{A(B)}$, repectively. However, $N^{A(B)}$ denotes the bosonic particles number. Furthermore, $U_0^{i}$ (i=A,B)  is the vacuum Rabi  frequency, while
	\begin{equation}
		|\eta_{A(B)}|=\sqrt{\frac{W \kappa_{A(B)}}{\hbar \omega_p^{A(B)}}}
	\end{equation}
	is the output power, and $\kappa_{A(B)}$ denotes decay rate associeted with cavity $A(B)$ and $W$ is the input laser power. The parameter $\zeta_{A(B)}$   describes the coupling strength parameter  characterizing the moving end mirror and intra-cavity field subsystems. The coupling parameter, namely $ \xi_m^{i}$ $(i=A, B)$ are associated to the coupling frequency of the field with the atom in each cavity.  Finally, let us suppose that the joint optomechanical cavities are pumped by two-mode squeezed light sources (SLS), with identical frequencies $\omega_{SL}$. Indeed, the SLS are distinguished by  the photon number and the two-photon correlations, namely $N(\omega_k)$ and $C(\omega_k)$. They are defined as \cite{k3},
	\begin{eqnarray}
		N(\omega_k)&=&\frac{\Lambda_1^2-\Lambda_2^2}{4} \big[ \frac{1}{(\omega_k-\omega_{SL})^2+\Lambda_2^2}-\frac{1}{(\omega_k-\omega_{SL})^2+\Lambda_1^2}\big],  \nonumber  \\
		C(\omega_k)&=&\frac{\Lambda_1^2-\Lambda_2^2}{4} \big[ \frac{1}{(\omega_k-\omega_{SL})^2+\Lambda_2^2}+\frac{1}{(\omega_k-\omega_{SL})^2+\Lambda_1^2}\big], 
	\end{eqnarray}
	
	where 
	\begin{eqnarray}
		\Lambda_1&=&\frac{1}{2}\pi_1- \pi_2,  \nonumber  \\
		\Lambda_1&=&\frac{1}{2}\pi_1+\pi_2,	 
	\end{eqnarray}
	
	where $\pi_1$ and $\pi_2$ denote the damping strength and amplification parameter associated to the squeezed light source. Without loss of generality, the quantities $N(\omega_k)$ and $ C(\omega_k)$ can be investigated independently of frequency. The ideal squeezed state implies maximum correlations squeezing to the case $M=\sqrt{N(N+1)}$ (for more details See Refs. \cite{k4,k5}).

	\section{Quantum Langevin equations \& Covariance matrix}
	
	We focus on the dynamics of the system in the presence of the interaction between the optomechanical cavities. We derive the coupled  Langevin equations associated to the operators $ \hat{a},  \hat{b},  \hat{p}_{A(B)},  \hat{q}_{A(B)}, \hat{P}_{A(B)}$ and $\hat{Q}_{A(B)}$ using the Hamiltonian defined in Eq.(\ref{eq1}). The Langevin equations are obtained as the following form 
	
	\begin{eqnarray}\label{Lang}
		\frac{d\hat{a}}{dt}&=&(i \tilde{\Delta}_A+i\zeta_A\hat{q}_A-i\xi_m^{A}\hat{Q}_A-\kappa_A)\hat{a}-\eta_A+\sqrt{2\kappa_A} \hat{c}_{in}-iJ\hat{b},  \nonumber  \\
		\frac{d\hat{b}}{dt}&=&(i \Delta_B^{'}+i\zeta_B\hat{q}_B-i\xi_m^{B}\hat{Q}_B-\kappa_B)\hat{b}-\eta_B+\sqrt{2\kappa_B} \hat{d}_{in}-iJ\hat{a},  \nonumber  \\
		\frac{d\hat{p}_A}{dt}&=&-\omega_m^{A} \hat{q}_A+\xi_A\hat{a}^{\dagger} \hat{a}-\gamma_m^{A} \hat{p}_A+\hat{I}_A(t), \nonumber  \\
		\frac{d\hat{p}_B}{dt}&=&-\omega_m^{B} \hat{q}_B+\xi_B\hat{b}^{\dagger} \hat{b}-\gamma_m^{B} \hat{p}_B+\hat{I}_B(t), \nonumber  \\
		\frac{d\hat{q}_A}{dt}&=&-\omega_m^{A} \hat{p}_A, \nonumber  \\
		\frac{d\hat{q}_B}{dt}&=&-\omega_m^{B} \hat{p}_B, \nonumber  \\
		\frac{d\hat{P}_A}{dt}&=&\Omega_{A} \hat{Q}_A+ \xi_m^{A} \hat{a}^{\dagger} \hat{a}-\gamma_{sm}^{A} \hat{P}_A+\hat{I}_{1m}^{A}(t),\nonumber  \\
		\frac{d\hat{P}_B}{dt}&=&\Omega_{B} \hat{Q}_B+ \xi_m^{B} \hat{b}^{\dagger} \hat{b}-\gamma_{sm}^{B} \hat{P}_B+\hat{I}_{1m}^{B}(t),\nonumber  \\
		\frac{d\hat{Q}_A}{dt}&=&-\Omega_{A} \hat{P}_A-\gamma_{sm}^{A} \hat{Q}_A+\hat{I}_{2m}^{A}(t),\nonumber  \\
		\frac{d\hat{Q}_B}{dt}&=&-\Omega_{B} \hat{P}_B-\gamma_{sm}^{B} \hat{Q}_B+\hat{I}_{2m}^{B}(t), 
	\end{eqnarray}
	
	where $\tilde{\Delta}_{A(B)}=\Delta_{A(B)}+\frac{U^{A(B)}N^{A(B)}}{2}$, and $\gamma_{m(sm)}^{A(B)}$ are the mechanical energy decays, while $\hat{I}_{A(B)}$  and  $\hat{I}_{1m(2m)}^{A(B)}$ are Brownian noise operators \cite{k6}. In general, they satisfy the following property
	
	\begin{equation}\label{t}
		\langle I_{A(B)}(t) I_{A(B)}(t^{'}) + I_{A(B)}(t^{'})  I_{A(B)}(t)\rangle/2= \gamma_{m}^{A(B)} (2 \bar{n}_{A(B)}+1)\delta(t-t^{'}), 
	\end{equation}
	such that $\bar{n}=(e^{\frac{\hbar \omega_m}{k_B T}}-1)^{-1}$	denotes the mean vibrational number, where, $k_B$ defines  Boltzmann constant,  $\omega_m$ is the mechanical resonator  frequency and $T$ is the temperature of the environment. Moreover, we have used the following properties:
	\begin{eqnarray}
		[\hat{a}, \hat{a}^{\dagger}]&=&\tilde{I}_d, \quad [\hat{b}, \hat{b}^{\dagger}]=\tilde{I}_d, \nonumber
	\end{eqnarray}
	\begin{eqnarray}
		[\hat{q}, \hat{p}]&=&i I_d, \quad [\hat{Q}, \hat{P}]=i I_d. \nonumber 
	\end{eqnarray}
	Finally, the input-output theory is also used via following formulas \cite{k7}
	\begin{eqnarray}
		\hat{a}(t)&=& \sqrt{2 \kappa_A} \hat{c}_{in}(t)-\kappa_A \hat{a}(t), \nonumber\\
		\hat{b}(t)&=&\sqrt{2 \kappa_B} \hat{d}_{in}(t)-\kappa_B \hat{b}(t).\nonumber
	\end{eqnarray}
	
The first and second Langevin equations in (\ref{Lang}) clearly show that the optical resonators of modes $\hat{a}$ and $\hat{b}$ are mathematically coupled via the coupling strength, $J$. Hence, based on the coupled  Langevin equations (\ref{Lang}), one can straightforwardly illustrate the covariance matrix of the proposed optomechanical system (for more details see Appendix A). In fact, the covariance matrix is has a $(12\times12)$-dimensional matrix. However, in order to investigate quantum correlation by means of Gaussian quantum discord,  we shall illustrate from the whole $(8\times8)$ covariance matrix  $C_v$ an eight $(4\times4)$ submatrices, namely $C_v^{(1)}, C_v^{(2)}$,....,$C_v^{(8)}$ obtained by taking trace operator into acount. The first block covariance matrix, namely $C_v^{(1)}$ of the elements kept to the set $\{1,2,3,4\}$ is related to the first intracavity photon-phonon, while the second block covariance matrix, namely $C_v^{(2)}$ of the elements kept to the set $\{5,6,7,8\}$ is attached to the second intracavity photon-phonon. The elements mechanical modes covariance matrix $C_v^{(3)}$ are kept to  set $\{1,2,5,6\}$, while the elements cavity modes covariance matrix, namely $C_v^{(4)}$are kept to  set $\{3, 4, 7, 8\}$. On the other hand, the block covariance matrices, namely $C_v^{(5)}$ and $C_v^{(6)}$ of elements kept to the set $\{1, 2, 9, 10\}$ and $\{5, 6, 11, 12\}$ examine the interaction between BEC with the first and second mechanical modes, respectively. Finally,  the block covariance matrices, namely $C_v^{(7)}$ and $C_v^{(8)}$ of elements kept to the set $\{3, 4, 9, 10\}$ and $\{7, 8, 11, 12\}$ examine the interaction between BEC with the first and second mechanical modes, respectively.

	\section{Gaussian quantum discord}
	
	Gaussian quantum states represents  the core of quantum information through continuous variables. However, the quantum  physical statement of quantum information is entanglement. Indeed, the entanglement  phenomenon forms a major resource in many tasks in secure information. Interestingly enough, continuous variable entanglement, i.e,  entangled Gaussian states  have been demonstrated as a worthy measure used to develop several  proposals; including cloning, teleportation quantum cryptography, etc. But,  entanglement still one kind of the so called quantum correlations. In fact, it have been shown that quantum correlations beyond entanglement can dominate the classical limits in a various proposed cases \cite{Adesso}. Hence, quantum correlations is recognized as a more general way to quantify the separability between quantum systems. In particular, a valuable class of continuous variables correlations measure is Gaussian quantum discord \cite{k11}.  Quantum discord is investigated as the difference between two expressions of mutual information. The last ones  quantumly analogs of  classically equivalent information.  For a bipartite Gaussian state which is described by its two-mode  covariance matrix $AB$ 
	
	\begin{equation}\label{hzt}
		\sigma=\begin{pmatrix}
			\sigma_1& \sigma_3 \\
			\sigma_3^{T}  & \sigma_2
		\end{pmatrix}.
	\end{equation}

	Here $\sigma_1$ and $\sigma_2$ are the covariance matrices attached to the sub-states of system $A$ and $B$, respectively. Moreover,  the matrix $\sigma_3$ characterizes  the correlations between the two subsystems, namely $A$ and $B$. Hence, the Guassian quantum discord is defined as  \cite{k12,k13}  
	\begin{equation}\label{res}
		\mathcal{D}=h(\sqrt{b_2})-h(\sqrt{s^{-}})-h(\sqrt{s^{+}})+h\big( \frac{\sqrt{b_1+2\sqrt{b_1b_2}+2I_3}}{1+2\sqrt{b_2}}\big), 
	\end{equation}
	where, 
	\begin{eqnarray}
		h(x)&=&(x+\frac{1}{2})\log(x+\frac{1}{2})-(x-\frac{1}{2})\log(x-\frac{1}{2}), \nonumber \\
		s^{\pm}	&=&\frac{1}{\sqrt{2}} \sqrt{Z\pm\sqrt{Z^2-4b_4}}, \nonumber \\
		Z&=&b_1+b_2+2I_3,  \nonumber \\
		b_1&=&\det \sigma_1, \nonumber \\
		b_2&=&\det \sigma_2, \nonumber \\
		b_3&=&\det \sigma_3, \nonumber \\
		b_4&=&\det \sigma.
	\end{eqnarray}
	where $\sigma_i$ ($i=1,2,3$) are defined in Eq.(\ref{hzt}). The quantities $b_j$ ($j=1,...,4$) are the so-called symplectic invariants which are still unchanged by all transformations  \cite{k13}, while $s^{\pm}$ define the symplectic eigenvalues of $\sigma$. As mentioned before, we shall investigate eight different block covariant matrices, namely $C_v^{(i)}$. Indeed, the symplectic eigenvalues of partial transpose of each submatrix. It takes the following compact form:
	\begin{equation}
		s^{\pm}=\frac{1}{\sqrt{2}} \big(f_1^{i}\pm\sqrt{f_1^{i^2}-4 f_2^{i}}\big)^{1/2}, 
	\end{equation}
	where $f_1^{i}=\det V_1^{i}+\det V_2^{i}-2\det V_3^{i}$ and $f_2^{i}=\det C_v^{i}$ $(i=1,2,3)$. The matrices $V_1^{i}, V_2^{i}$ and $V_3^{i}$ being the $(2\times2)$ block matrices of the whole ($4\times 4$) Guassian states $C_v^{i}$ as
	\begin{equation}\label{ft}
		C_v^{i}=\begin{pmatrix}
			V_1^{i} & V_3^{i}  \\
			V_3^{i^{T}}   & V_2^{i}
		\end{pmatrix}.
	\end{equation} 
	Additionally, for any bipartite Gaussian state defined  by its two-mode  covariance matrix is entangled when the Gaussian quantum discord in Eq.(	\ref{res}) satisfies $\mathcal{D}>1$, but this state is unentangled or entangled when $0\leq\mathcal{D}\leq1$. The Gaussian quantum discord has been experimentally demonstrated  \cite{k14,k15}. Within this framework, we are able to quantify the distribution of Gaussian quantum discord between various subsystems of the current study, namely the first intracavity photon-phonon, the second intracavity photon-phonon  and the cavity modes subsystems.
	
	\subsection{Gaussian quantum discord versus $\Delta/\omega_m$ for the first and  second intracavity photon-phonon system}
	
	Before proceeding, it is worth noting that in the plots below, we have chosen some of the parameters regime encoded in the covariance matrix that are close to the topical experiments (See those given in Refs.\cite{k16,k17,k18}) which guarantee the  stability of the system (See Appendix B). Besides, one can  estimate different values of the other parameters, namely the detuning parameter $\Delta$ and the mean number of photons, namely $n$ and $N$, to quantify the Gaussian quantum discord. Moreover, for the sake of simplicity, we assume that both cavities are identical. In this inspiration, we put $\omega_m^{A}=\omega_m^{B}=\omega_m, \kappa_A=\kappa_B=\kappa, \gamma_m^{A}=\gamma_m^{B}=\gamma_m,T_A=T_B=T,  \Delta_A=\Delta_B=\Delta$, $\Omega_A=\Omega_B=\Omega$, $\gamma_{sm}^{A}=\gamma_{sm}^{B}=\gamma_{sm}$ and $\bar{n}_A=\bar{n}_B=\bar{n}$. Indeed, since the cavities are perfectly identical, then in our  numerical results we have obtained the same covariance matrices for the first intracavity photon-phonon and  second intracavity photon-phonon. Therefore, in this case, we shall obtain the same behavior for Gaussian quantum discord. However, regardless of the intracavity photon-phonon system, the cavity modes, mechanical resonator modes, BEC-photon and BEC-phonon systems can be also used to quantify the Gaussian  quantum discord in our model. \\

	\begin{figure}[h!]
		\begin{center}
			\includegraphics[scale=.64]{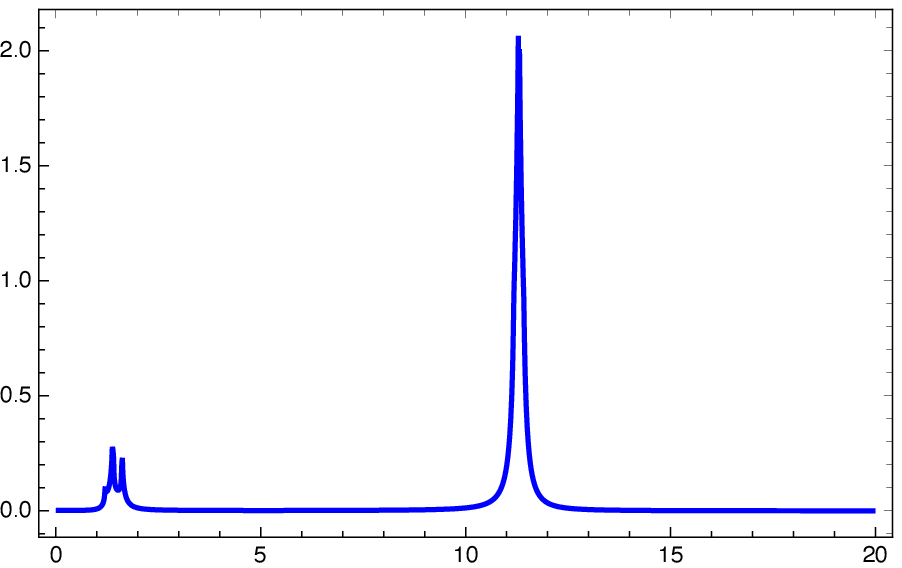}
			\put(-160,110){$\mathcal{D}$}
			\put(-90,-10){$\Delta/\omega_m$}
			\put(-90,110){\textbf{(a)}}
			~~~~~~~~\quad
			\includegraphics[scale=.64]{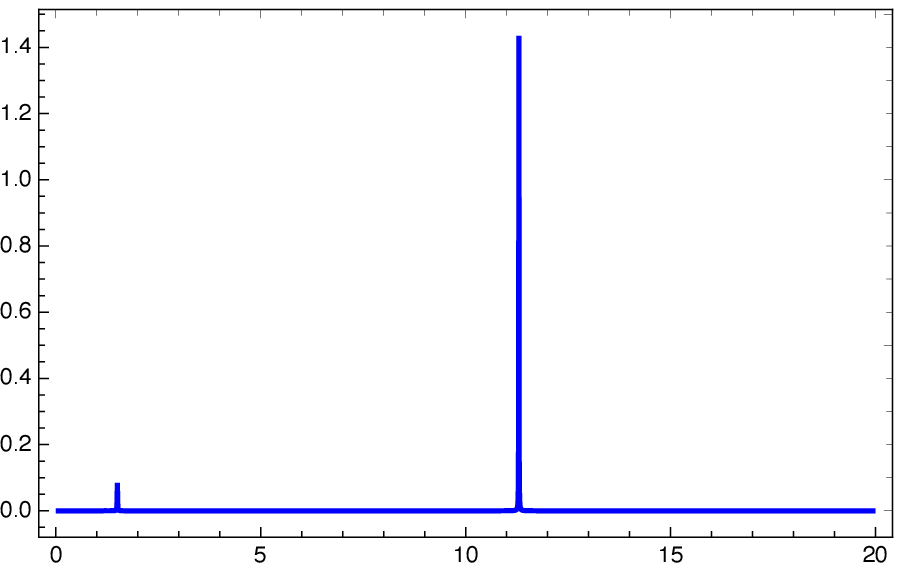}
			\put(-160,110){$\mathcal{D}$}
			\put(-90,-10){$\Delta/\omega_m$}
			\put(-90,110){\textbf{(b)}}\\
			\includegraphics[scale=.64]{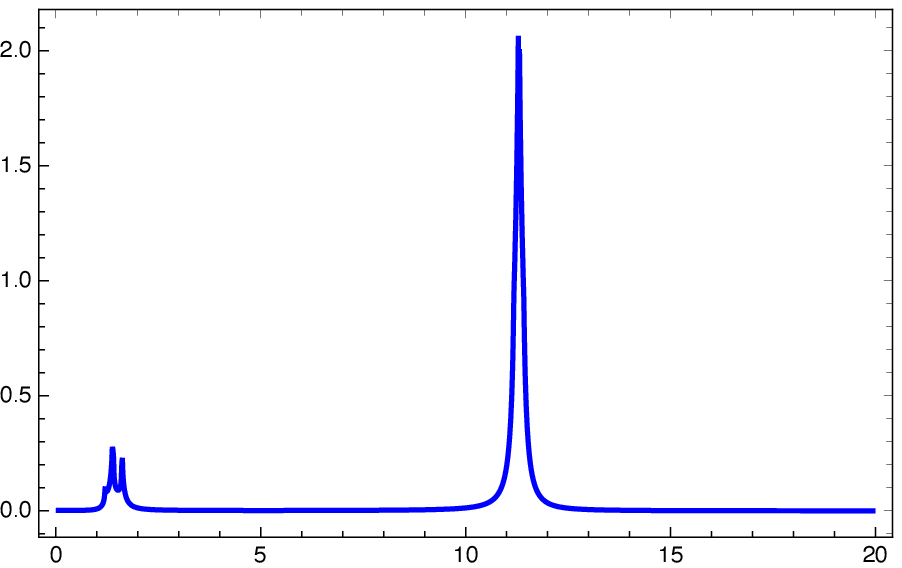}
			\put(-160,110){$\mathcal{D}$}
			\put(-90,-10){$\Delta/\omega_m$}
			\put(-90,110){\textbf{(c)}}
			~~~~~~~~\quad
			\includegraphics[scale=.64]{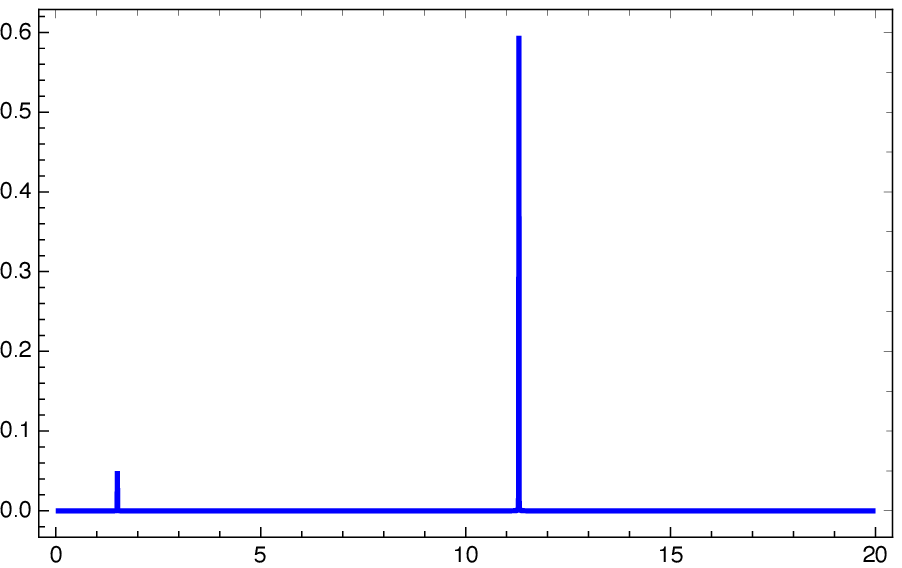}
			\put(-160,110){$\mathcal{D}$}
			\put(-90,-10){$\Delta/\omega_m$}
			\put(-90,110){\textbf{(d)}}
			\caption{Dynamics of Gaussian  quantum discord  versus the normalized detuning parameter $\Delta/\omega_m$, for the first and  second intracavity photon-phonon system. We set,  $\kappa=14 \times 2 \pi $ MHz, $\mu^{A}=\mu^{B}=S_A=S_B=8 \times 2 \pi$ MHz, $\gamma_m=\gamma_{sm}=100 \times 2 \pi$ MHz, $\Omega=10 \times 2 \pi$. Moreover, (a)  $\bar{n}=N=0$, $J=1Hz$, (b) $\bar{n}=836, N=0$, $J=1Hz$, (c) $\bar{n}=N=0$, $J=0.5Hz$ and (d) $\bar{n}=14642, N=0.1$, $J=1Hz$.}
			\label{f1}
		\end{center}
	\end{figure}
	
	The plots in Fig. (\ref{f1}) display that the Gaussian quantum discord against the normalized detuning parameter, namely $\Delta/\omega_m$ by consideration of various situations. Indeed, in Fig. (\ref{f1}a) we set $\bar{n}=N=0$ and $J=1$. The plot shows that  the Gaussian  quantum discord for the first and second intracavity photon-phonon systems disappears when the normalized detuning parameter vanishes. As  $\Delta/\omega_m$ takes small numbers, the Gaussian quantum discord behaves with small bounds until  maximized around $\Delta/\omega_m=11.5$. A remarkable decreasing appears fast for large numbers of the  parameter $\Delta/\omega_m$. Indeed, once the normalized detuning parameter increases, the Gaussian discord decreases monotonically until occuring the minimum bounds. On the other hand,  the Gaussian discord exceeds one, only when $\Delta/\omega_m=11.5$ which means that the state is entangled only for this critical value. While, for $\Delta/\omega_m\neq11.5$ the covariance matrix can be entangled or unentangled since $\mathcal{D}<1$. In Fig. (\ref{f1}a) we investigate the influence of increasing the parameter $\bar{n}$. The obtained plot shows the smme behaviour but now it is compacted with different upper bound. Indeed, it is clear that the Guassian quantum discord decreases comparing to Fig. (\ref{f1}a) but the state still always entangled for the critical value of the normalized detuning parameter, namely $\Delta/\omega_m=11.5$. Now, let examine the impact of small value of the coupling $J$ by keeping always the same initial settings as in Fig. (\ref{f1}a). A similar iour is obtained which means that in this case the parameter $J$  has not a clear effect on the dynamics of Guassian quantum discord . The plot in Fig. (\ref{f1}c) shows the dynamics of quantum correlation by choosing robust values of $\bar{n}$ and $N$. It is clear that in this case the state is unentangled for all $\Delta/\omega_m$.\\

	\subsection{Gaussian quantum discord versus $\Delta/\omega_m$ for the two mechanical resonator modes system}

	\begin{figure}[h!]
	\begin{center}
		\includegraphics[scale=.64]{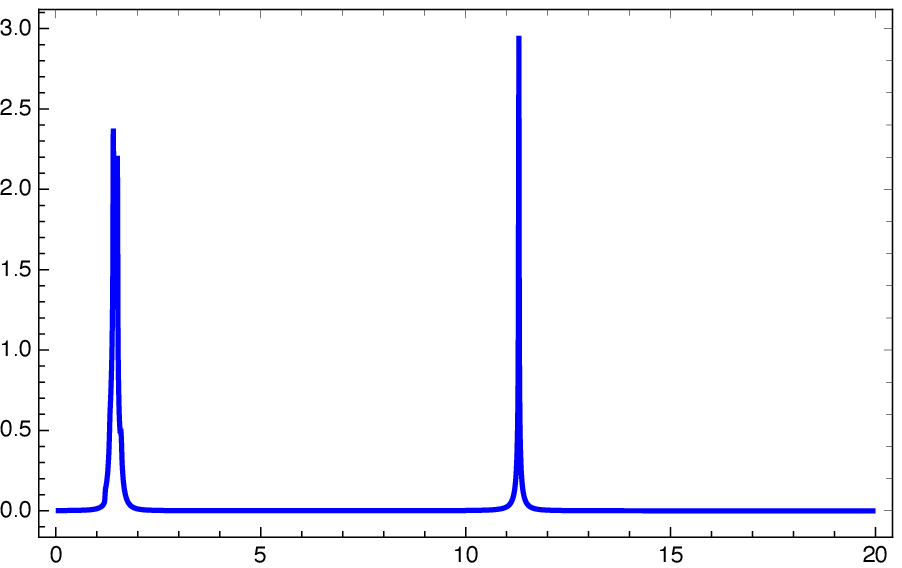}
	\put(-160,110){$\mathcal{D}$}
	\put(-90,-10){$\Delta/\omega_m$}
	\put(-90,110){\textbf{(a)}}
	~~~~~~~~\quad
	\includegraphics[scale=.64]{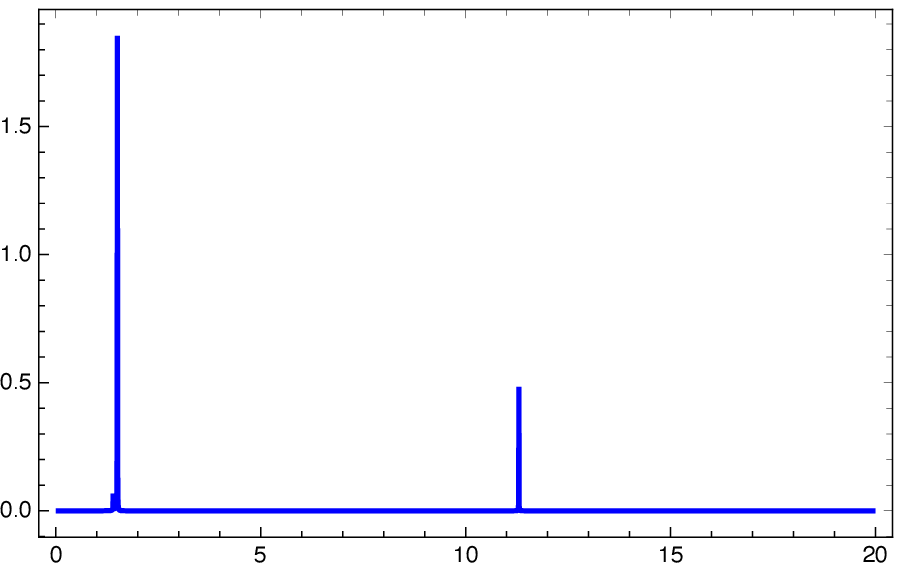}
	\put(-160,110){$\mathcal{D}$}
	\put(-90,-10){$\Delta/\omega_m$}
	\put(-90,110){\textbf{(b)}}\\
\includegraphics[scale=.64]{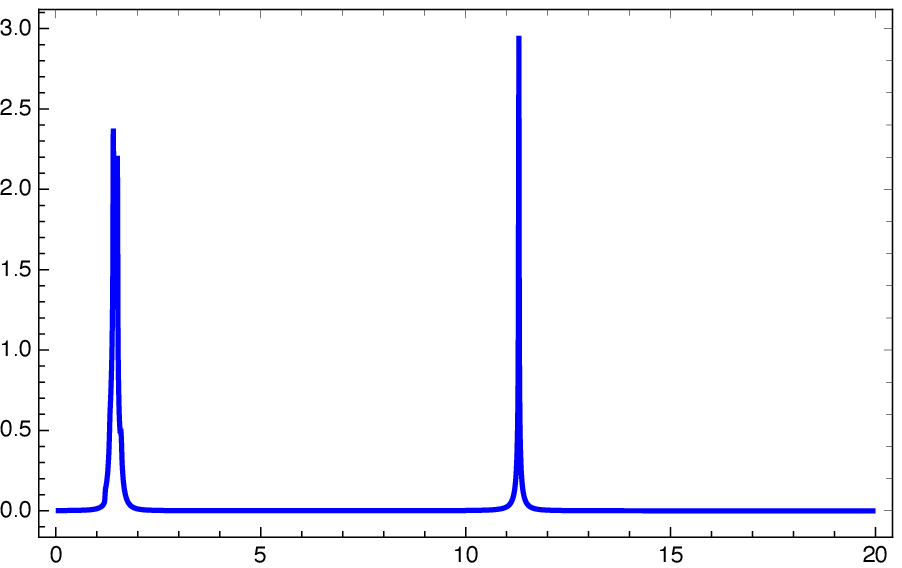}
\put(-160,110){$\mathcal{D}$}
\put(-90,-10){$\Delta/\omega_m$}
\put(-90,110){\textbf{(c)}}
~~~~~~~~\quad
\includegraphics[scale=.64]{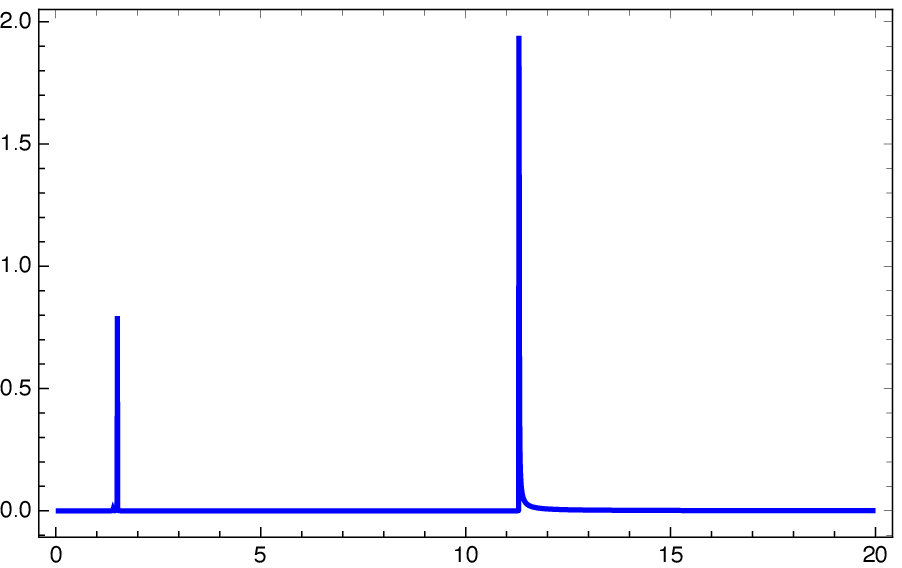}
\put(-160,110){$\mathcal{D}$}
\put(-90,-10){$\Delta/\omega_m$}
\put(-90,110){\textbf{(d)}}
		\caption{Dynamics of Gaussian  quantum discord  versus the normalized detuning parameter $\Delta/\omega_m$, for the two mechanical resonator modes system.  We set,  $\kappa=14 \times 2 \pi $ MHz, $\mu^{A}=\mu^{B}=S_A=S_B=8 \times 2 \pi$ MHz, $\gamma_m=\gamma_{sm}=100 \times 2 \pi$ MHz, $\Omega=10 \times 2 \pi$. Moreover, (a)  $\bar{n}=N=0$, $J=1Hz$, (b) $\bar{n}=836, N=0$, $J=1Hz$, (c) $\bar{n}=N=0$, $J=0.5Hz$ and (d) $\bar{n}=14642, N=0.1$, $J=1Hz$.}
		\label{f2}
	\end{center}
\end{figure}
	
In Fig.(\ref{f2}), we numerically plot the dynamics of Gaussian quantum discord  against the normalized  detuning parameter, namely $\Delta_m/\omega_m$ for the two mechanical resonator modes system. Initially, the Guassian quantum discord is initially vanished by considering various initial settings of coupled hybrid optomechanical cavities. After a short interval of interaction time $t$ the quantum correlation shows the first peak of $\mathcal{D}>1$ (correlated cavity-cavity state) except for Fig.(\ref{f2}d) when robust values of the mean numbers of photons $\bar{n}$ and $N$ are considered.  For $t>2$ the Gaussian quantum discord completely vanishes and then increases dramatically with the increasing of the normalized detuning parameter until reached the maximum value around $\Delta/\omega_m \approx 11$. Moreover, it is clear that the Gaussian quantum discord still non-sensitive to changes in the normalized detuning parameter, for large values of the interaction time parameter, namely $t$. From Figs.(\ref{f1}) and .(\ref{f2}) one can conclude that the quantum correlations are enhanced using two mechanical resonator modes system. \\

		\subsection{Gaussian quantum discord versus $\Delta/\omega_m$ for the cavity modes system}
		
		\begin{figure}[h!]
		\begin{center}
			\includegraphics[scale=.64]{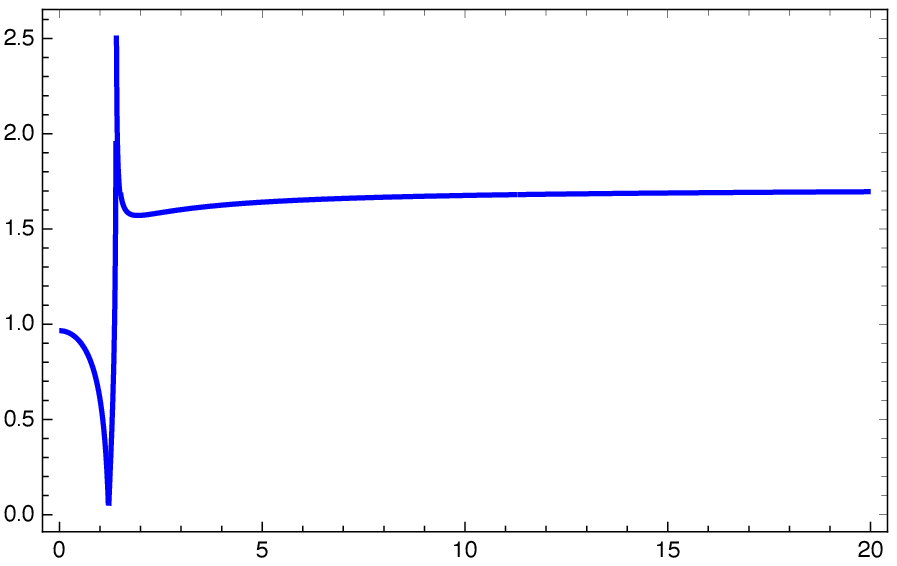}
		\put(-160,110){$\mathcal{D}$}
		\put(-90,-10){$\Delta/\omega_m$}
		\put(-90,110){\textbf{(a)}}
		~~~~~~~~\quad
		\includegraphics[scale=.64]{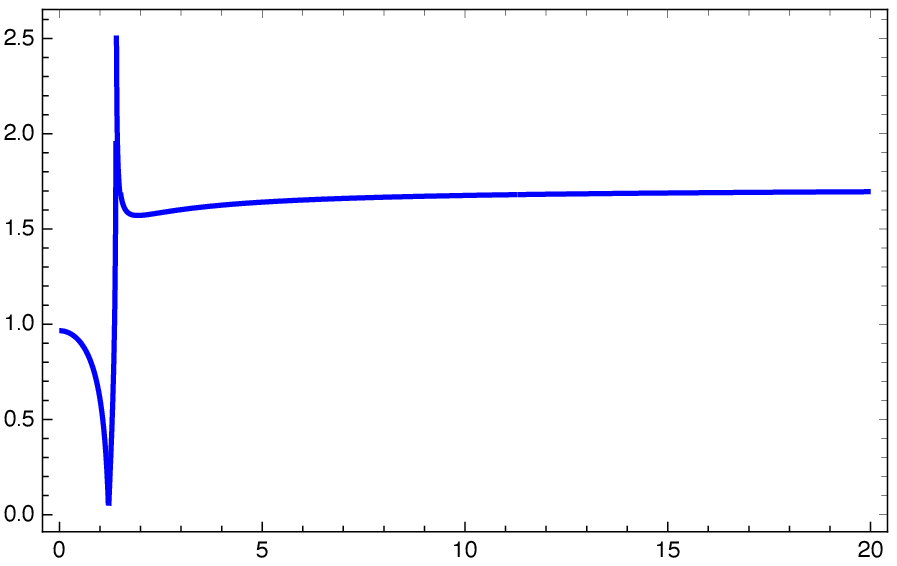}
		\put(-160,110){$\mathcal{D}$}
		\put(-90,-10){$\Delta/\omega_m$}
		\put(-90,110){\textbf{(b)}}\\
		\includegraphics[scale=.64]{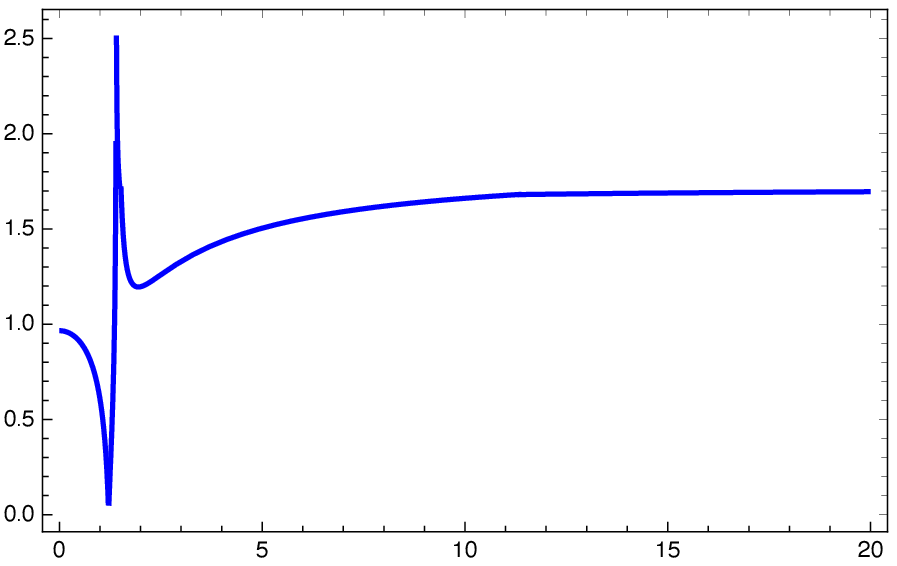}
		\put(-160,110){$\mathcal{D}$}
		\put(-90,-10){$\Delta/\omega_m$}
		\put(-90,110){\textbf{(c)}}
		~~~~~~~~\quad
		\includegraphics[scale=.64]{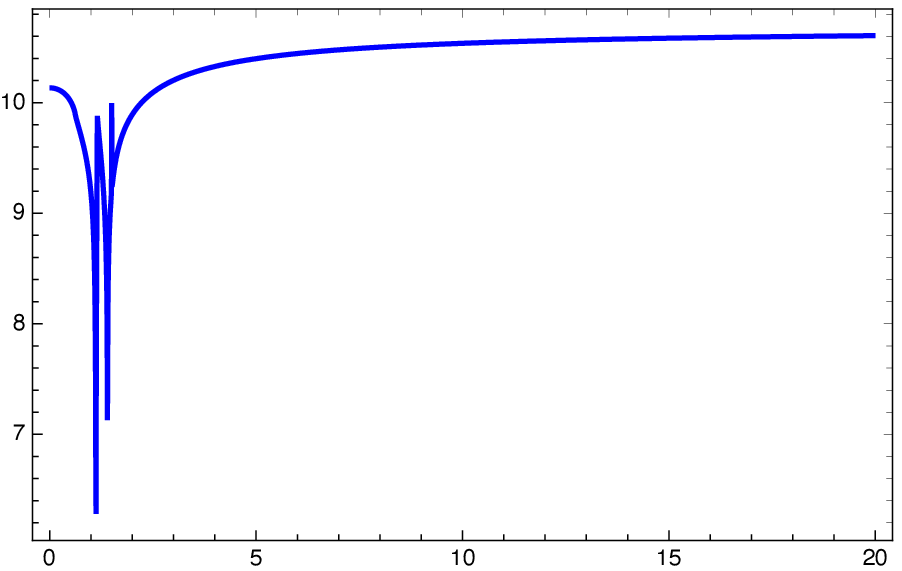}
		\put(-160,110){$\mathcal{D}$}
		\put(-90,-10){$\Delta/\omega_m$}
		\put(-90,110){\textbf{(d)}}
		\caption{Dynamics of Gaussian  quantum discord  versus the normalized detuning parameter $\Delta/\omega_m$, for  the cavity modes system.  We set,  $\kappa=14 \times 2 \pi $ MHz, $\mu^{A}=\mu^{B}=S_A=S_B=8 \times 2 \pi$ MHz, $\gamma_m=\gamma_{sm}=100 \times 2 \pi$ MHz, $\Omega=10 \times 2 \pi$. Moreover, (a)  $\bar{n}=N=0$, $J=1Hz$, (b) $\bar{n}=836, N=0$, $J=1Hz$, (c) $\bar{n}=N=0$, $J=0.5Hz$ and (d) $\bar{n}=14642, N=0.1$, $J=1Hz$.}
			\label{f3}
		\end{center}
	\end{figure}

Now, let us investigate the quantum correlation of bipartite continuous variable system in terms of the Gaussian quantum discord using intracavity cavity modes system against the normalized detuning, namely $\Delta/\omega_m$ by varying various parameters encoded in the sub-covariance matrix $C_v^{4}$ obtained from Eq.(\ref{Lang}). Indeed, from Figs.(\ref{f3}a)-(\ref{f3}c) it is obvious that we recover the same behavior of  the Gaussian quantum discord of the cavity modes system for different values of the coupling constant $J$ and the mean number of photons $\bar{n}$. It is clear that the Guassian quantum discord for $\Delta/\omega_m=0$ takes the unity value, while as the normalized detuning parameter increases the quantum discord  decreases fast which means that the covariance matrix is either entangled or unentangled since the Gaussian quantum discord indicates an intermediate value between zero and unity. Furthermore, the Gaussian quantum discord increases to owns a maximum pick for small numbers of  the normalized detuning parameter, arround a critical value, namely $\Delta/\omega_m=1$. When the normalized detuning parameter  is larger than this critical value, the  Gaussian quantum discord remains constant even for robust values of  $\Delta/\omega_m$. In Fig.(\ref{f3}d), we consider the non-zeros of the mean numbers of photons, that is, $\bar{n}$ and $N$. For this particular case, we clearly observe that the covariance matrix for the cavity modes system is always correlated for all considered values of $\Delta/\omega_m$ since $\mathcal{D}>1$. Hence, one can conclude that by controlling various physical parameter of the system ($\bar{n}, N$ and $J$) we can enhance considerabely the amount of quantum correlations by means of Gaussian quantum discord.\\

		\subsection{Gaussian quantum discord versus $\Delta/\omega_m$ for BEC-first mechanical mode system}
	
		\begin{figure}[h!]
		\begin{center}
			\includegraphics[scale=.64]{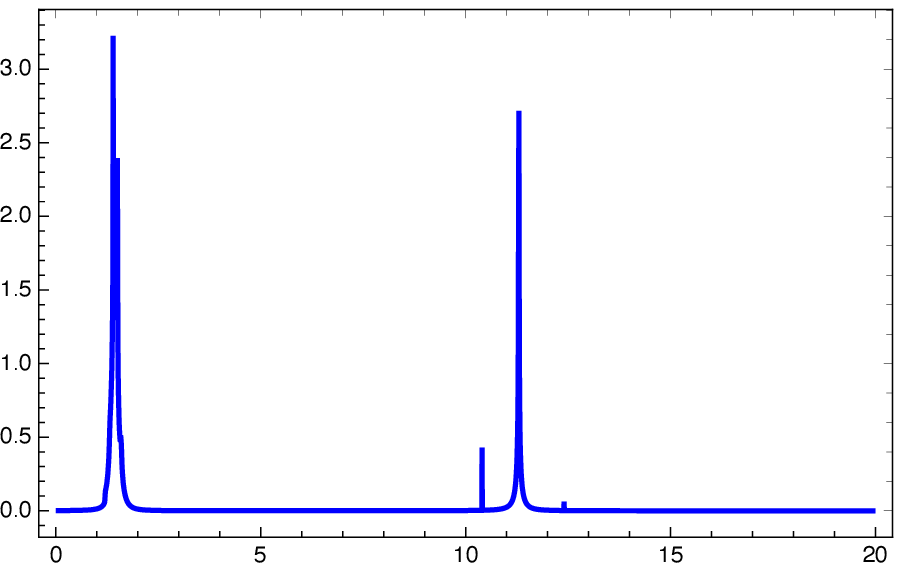}
		\put(-160,110){$\mathcal{D}$}
		\put(-90,-10){$\Delta/\omega_m$}
		\put(-90,110){\textbf{(a)}}
		~~~~~~~~\quad
		\includegraphics[scale=.64]{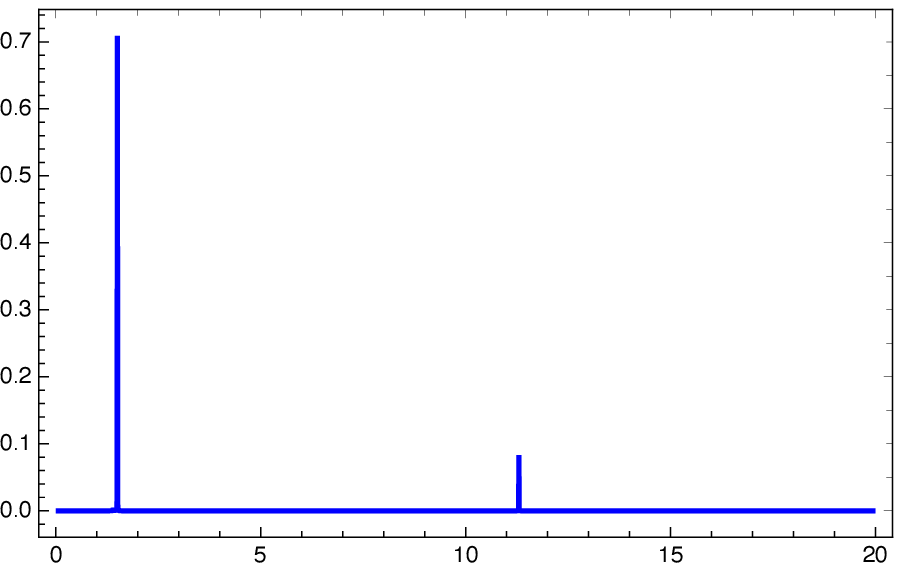}
		\put(-160,110){$\mathcal{D}$}
		\put(-90,-10){$\Delta/\omega_m$}
		\put(-90,110){\textbf{(b)}}\\
		\includegraphics[scale=.64]{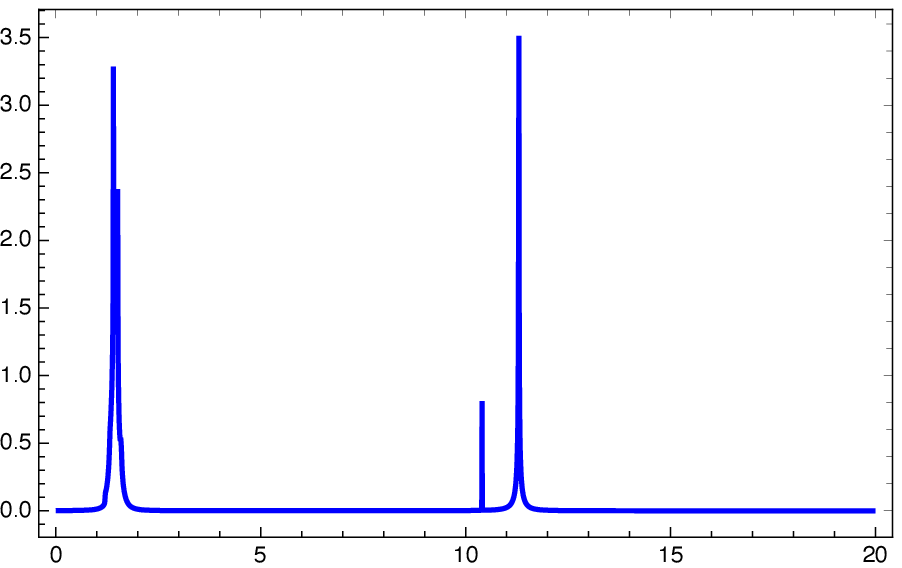}
		\put(-160,110){$\mathcal{D}$}
		\put(-90,-10){$\Delta/\omega_m$}
		\put(-90,110){\textbf{(c)}}
		~~~~~~~~\quad
		\includegraphics[scale=.64]{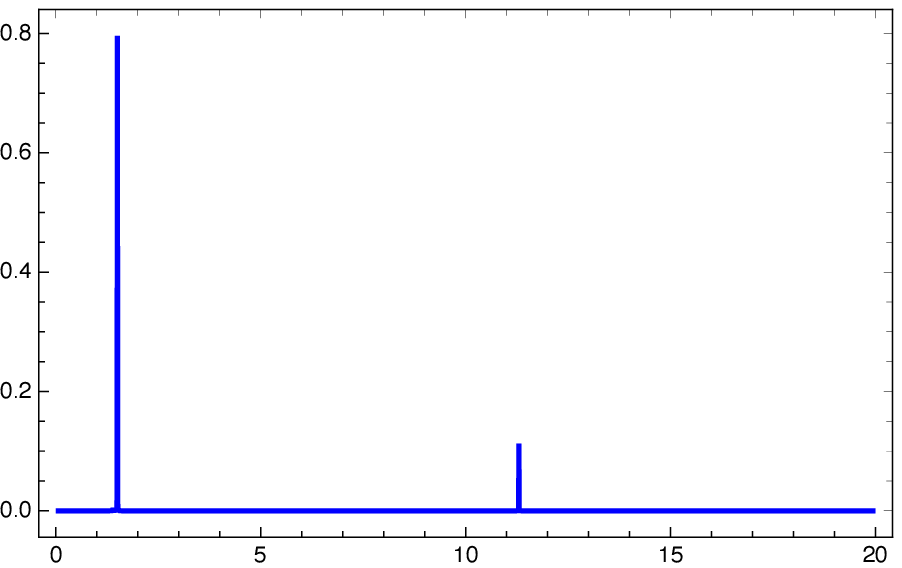}
		\put(-160,110){$\mathcal{D}$}
		\put(-90,-10){$\Delta/\omega_m$}
		\put(-90,110){\textbf{(d)}}
		\caption{Dynamics of Gaussian  quantum discord  versus the normalized detuning parameter $\Delta/\omega_m$, for BEC-first mechanical mode system.  We set,  $\kappa=14 \times 2 \pi $ MHz, $\mu^{A}=\mu^{B}=S_A=S_B=8 \times 2 \pi$ MHz, $\gamma_m=\gamma_{sm}=100 \times 2 \pi$ MHz, $\Omega=10 \times 2 \pi$. Moreover, (a)  $\bar{n}=N=0$, $J=1Hz$, (b) $\bar{n}=836, N=0$, $J=1Hz$, (c) $\bar{n}=N=0$, $J=0.5Hz$ and (d) $\bar{n}=14642, N=0.1$, $J=1Hz$.}
			\label{f4}
		\end{center}
	\end{figure}

We devote this subsection to examining the analytical findings of Gaussian quantum discord against the the normalized detuning parameter, namely $\Delta/\omega_m$ by considering the covariance matrix of the interacted system of BEC with the first mechanical mode. Indeed, in Fig.(\ref{f4}) the detuning parameter is used to quantify the non-classical correlation by means of Gaussian quantum discord. Obviously, for Figs.(\ref{f4}a) and (\ref{f4}c), we clearly see that the covariance matrix is entangled for $\Delta/\omega_m=1.5$ and $\Delta/\omega_m=11.5$. On the other hand, we assume that $\bar{n}=836, N=0$, $J=1Hz$ and $\bar{n}=14642, N=0.1$, $J=1Hz$, in Figs.(\ref{f4}b) and (\ref{f4}d), respectively. In this case  we clearly see that the Gaussian quantum discord is destroyed from the system s either entangled or unentangled since the Gaussian quantum discord indicates never exceeds the unity.

	\subsection{Gaussian quantum discord versus $\Delta/\omega_m$ for BEC with second mechanical mode}
	
		\begin{figure}[h!]
		\begin{center}
			\includegraphics[scale=.64]{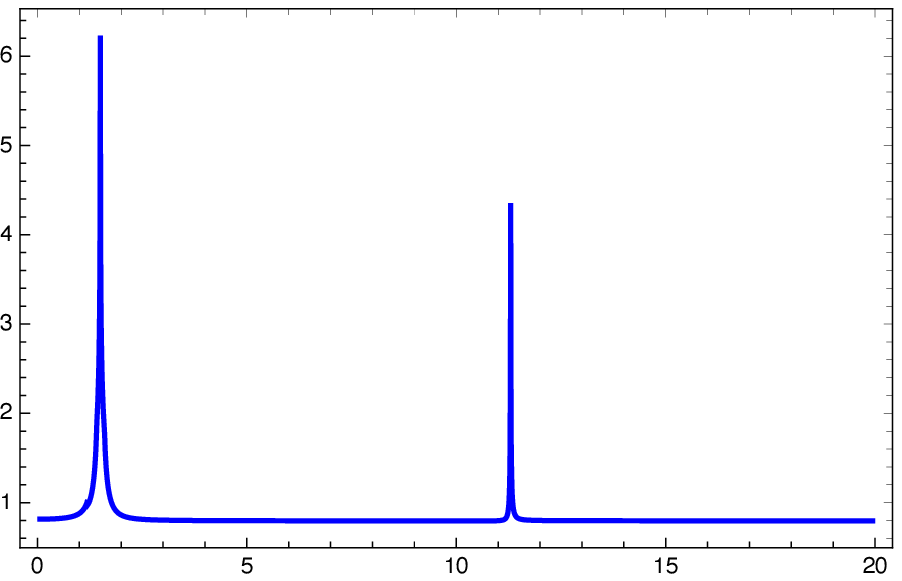}
		\put(-160,110){$\mathcal{D}$}
		\put(-90,-10){$\Delta/\omega_m$}
		\put(-90,110){\textbf{(a)}}
		~~~~~~~~\quad
		\includegraphics[scale=.64]{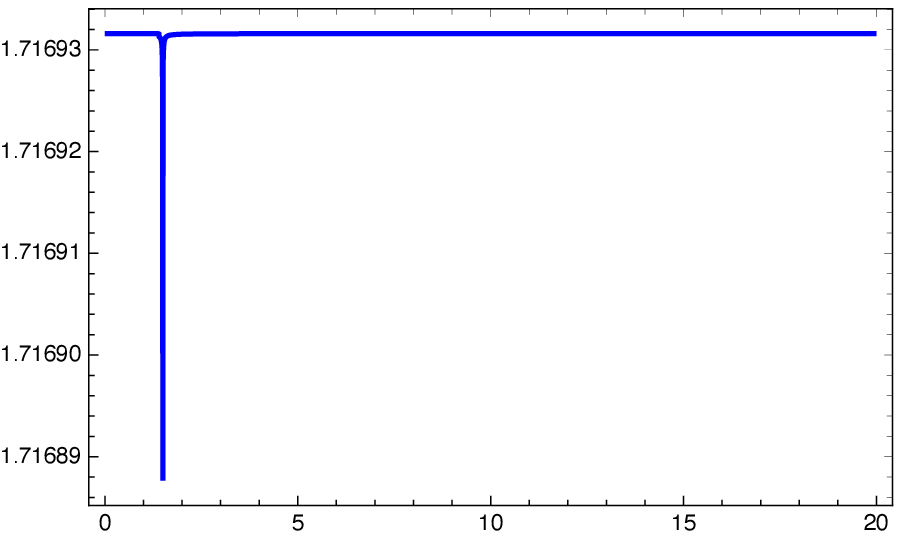}
		\put(-160,110){$\mathcal{D}$}
		\put(-90,-10){$\Delta/\omega_m$}
		\put(-90,110){\textbf{(b)}}\\
		\includegraphics[scale=.64]{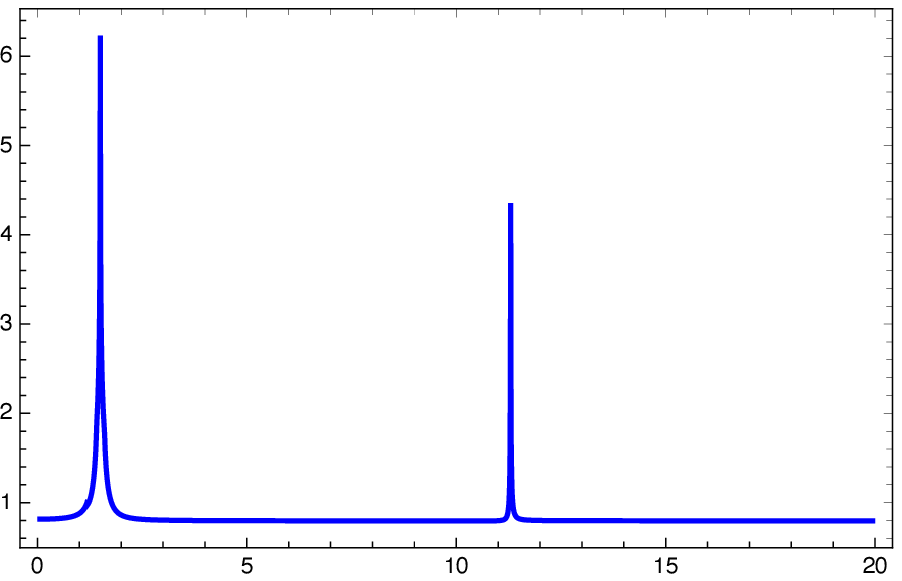}
		\put(-160,110){$\mathcal{D}$}
		\put(-90,-10){$\Delta/\omega_m$}
		\put(-90,110){\textbf{(c)}}
		~~~~~~~~\quad
		\includegraphics[scale=.64]{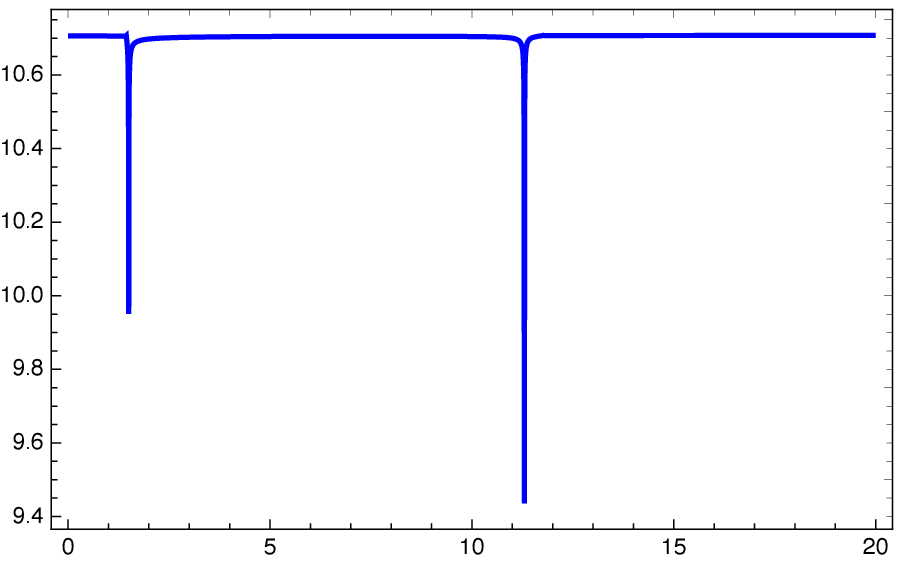}
		\put(-160,110){$\mathcal{D}$}
		\put(-90,-10){$\Delta/\omega_m$}
		\put(-90,110){\textbf{(d)}}
		\caption{Dynamics of Gaussian  quantum discord  versus the normalized detuning parameter $\Delta/\omega_m$, for BEC-second mechanical mode system.  We set,  $\kappa=14 \times 2 \pi $ MHz, $\mu^{A}=\mu^{B}=S_A=S_B=8 \times 2 \pi$ MHz, $\gamma_m=\gamma_{sm}=100 \times 2 \pi$ MHz, $\Omega=10 \times 2 \pi$. Moreover, (a)  $\bar{n}=N=0$, $J=1Hz$, (b) $\bar{n}=836, N=0$, $J=1Hz$, (c) $\bar{n}=N=0$, $J=0.5Hz$ and (d) $\bar{n}=14642, N=0.1$, $J=1Hz$.}
			\label{f5}
		\end{center}
	\end{figure}

For the sake of completeness, let’s now investigate the quantum correlation of bipartite continuous variable of BEC-second mechanical mode system in terms of the Gaussian quantum discord. When considering the zero-values of the mean number of photons, namely $\bar{n}=N=0$, we examine the quantum discord $\mathcal{D}$ either by choosing $J=1Hz$ (Fig.(\ref{f5}a)) or $J=0.5Hz$ (Fig.(\ref{f5}c)). A similar behaviours are obtained  where the sub-covariance matrix $C_v^{(6)}$ is always correlated for the critical values $\Delta/\omega_m=1.5$ and $\Delta/\omega_m=11.5$. For the rest of interval, the state is either entangled or separable ($\mathcal{D}<1$) . Furthermore, by raising the mean numbers of photons of the optical modes and SLS, namely $\bar{n}$ and $N$, respectively (see Figs.(\ref{f5}b) and Fig.(\ref{f5}d)), we observe that the covariance matrix described the interaction between BEC and the mechanical mode of the second optomechanical mode is always correlated since the Gaussian quantum discord indicates the amplitudes lager than unity. \\

\subsection{Gaussian quantum discord versus $\Delta/\omega_m$ for BEC with first optical mode}

	\begin{figure}[h!]
	\begin{center}
		\includegraphics[scale=.64]{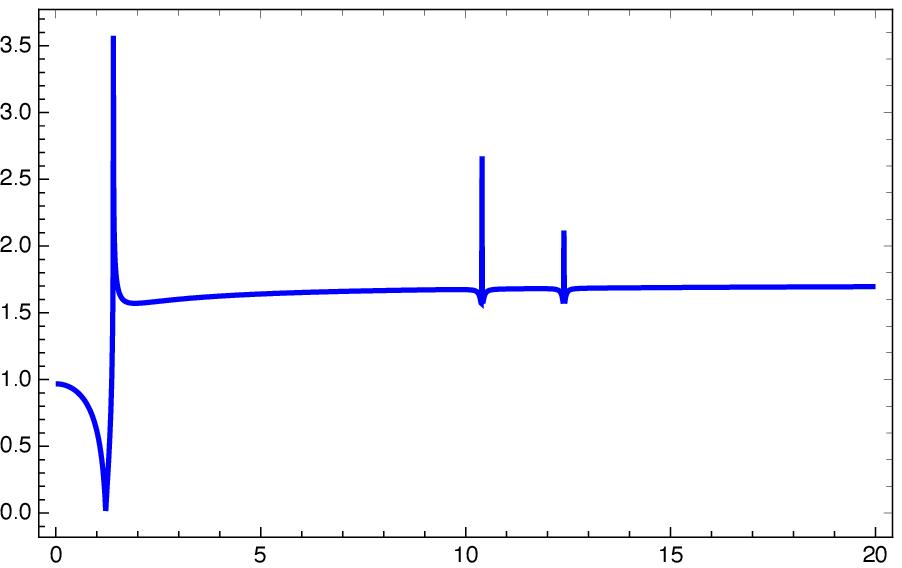}
	\put(-160,110){$\mathcal{D}$}
	\put(-90,-10){$\Delta/\omega_m$}
	\put(-90,110){\textbf{(a)}}
	~~~~~~~~\quad
	\includegraphics[scale=.64]{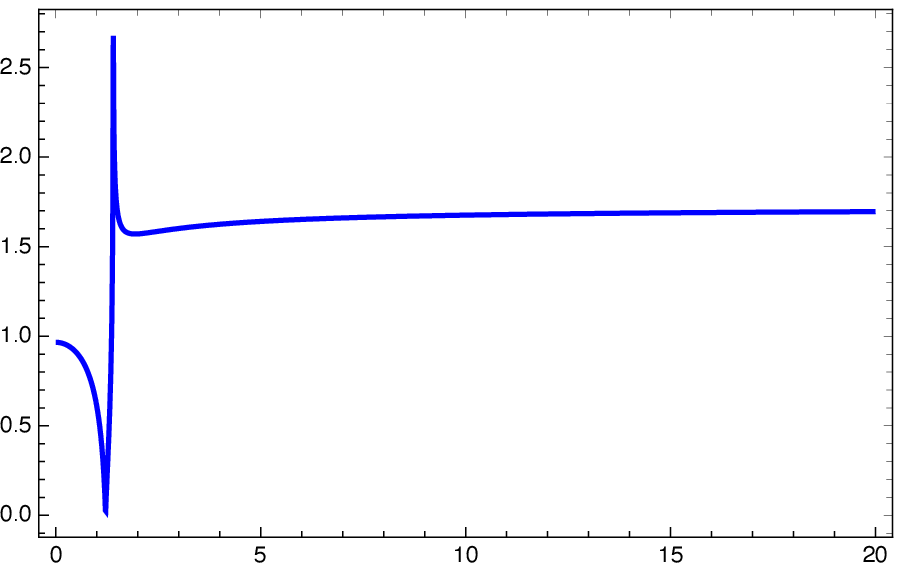}
	\put(-160,110){$\mathcal{D}$}
	\put(-90,-10){$\Delta/\omega_m$}
	\put(-90,110){\textbf{(b)}}\\
	\includegraphics[scale=.64]{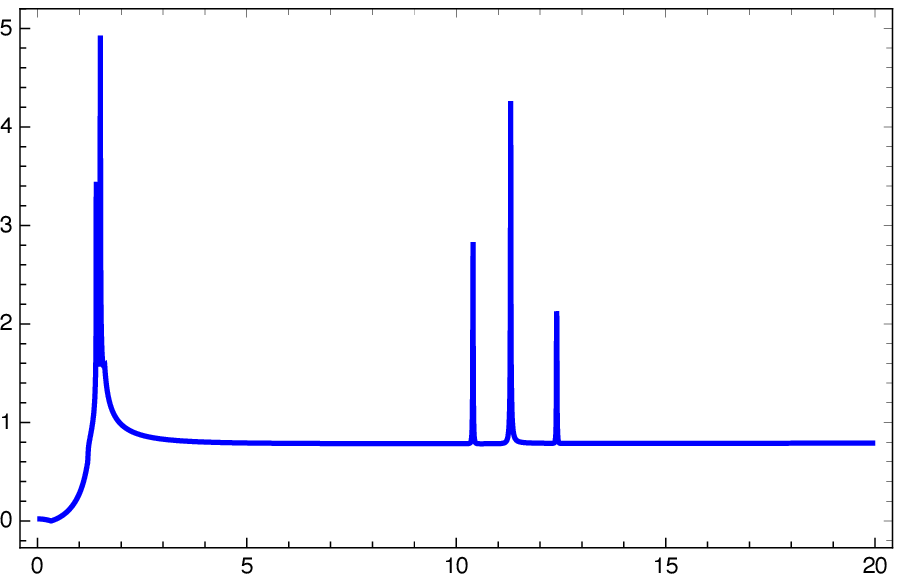}
	\put(-160,110){$\mathcal{D}$}
	\put(-90,-10){$\Delta/\omega_m$}
	\put(-90,110){\textbf{(c)}}
	~~~~~~~~\quad
	\includegraphics[scale=.64]{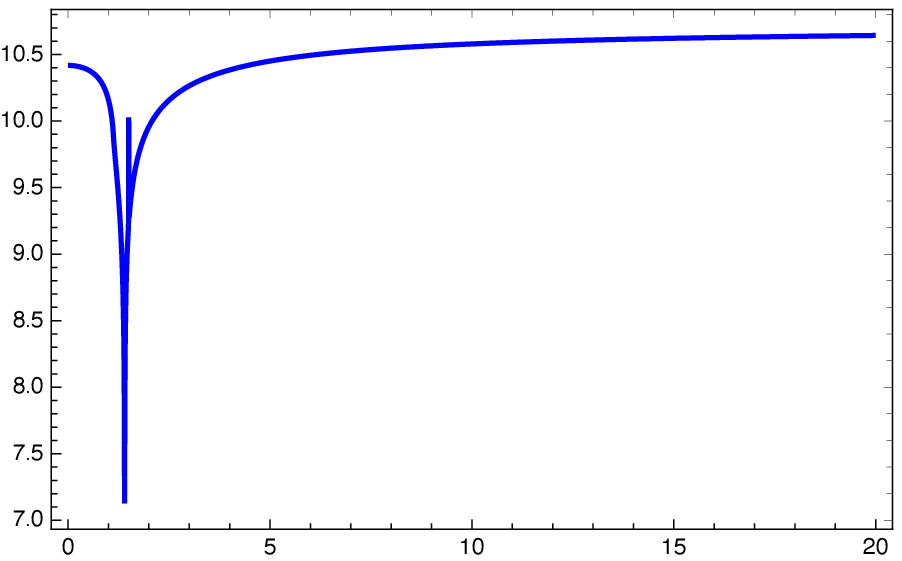}
	\put(-160,110){$\mathcal{D}$}
	\put(-90,-10){$\Delta/\omega_m$}
	\put(-90,110){\textbf{(d)}}
    \caption{Dynamics of Gaussian  quantum discord  versus the normalized detuning parameter $\Delta/\omega_m$, for BEC-first optical mode system.  We set,  $\kappa=14 \times 2 \pi $ MHz, $\mu^{A}=\mu^{B}=S_A=S_B=8 \times 2 \pi$ MHz, $\gamma_m=\gamma_{sm}=100 \times 2 \pi$ MHz, $\Omega=10 \times 2 \pi$. Moreover, (a)  $\bar{n}=N=0$, $J=1Hz$, (b) $\bar{n}=836, N=0$, $J=1Hz$, (c) $\bar{n}=N=0$, $J=0.5Hz$ and (d) $\bar{n}=14642, N=0.1$, $J=1Hz$.}
		\label{f6}
	\end{center}
\end{figure}

Now, let investigate the quantum correlation of bipartite continuous variable system in terms of the Gaussian quantum discord of the intracted BEC and optical inside the first optomechanical cavity against the normalized detuning parameter, namely $\Delta/\omega_m$ by varying various parameters encoded in the sub-covariance matrices $C_v^{7}$. Indeed, one can see from Figs.(\ref{f6}a) and (\ref{f6}c) that the Gaussian quantum discord is not vanished. As we increase gradually the normalized detuning number, we clearly see that the amount of quantum correlation fluctuate between its maximum and minimum bounds, while it remain remain constant for some specific intervales of $\Delta/\omega_m$. Again, by examining the large numbers of the means numbers of photons $\bar{n}$ and $N$ in Figs.(\ref{f6}b) and (\ref{f6}d) the behaviours exibits some few oscillations for small values of the normalized detuning parameter while the Guassian quantum discord remains constant for the remaining values of $\Delta/\omega_m$. Furthermore, for some intervals of detuning parameter, we clearly observe that the covariance matrix is either entangled or unentangled since the Gaussian quantum discord indicates an intermediate value between zero and unity. However for some critical values of  $\Delta/\omega_m$ the state described the BEC-first optical mode system is correlated. \\

\subsection{Gaussian quantum discord versus $\Delta/\omega_m$ for BEC with second optical mode}

	\begin{figure}[h!]
	\begin{center}
		\includegraphics[scale=.64]{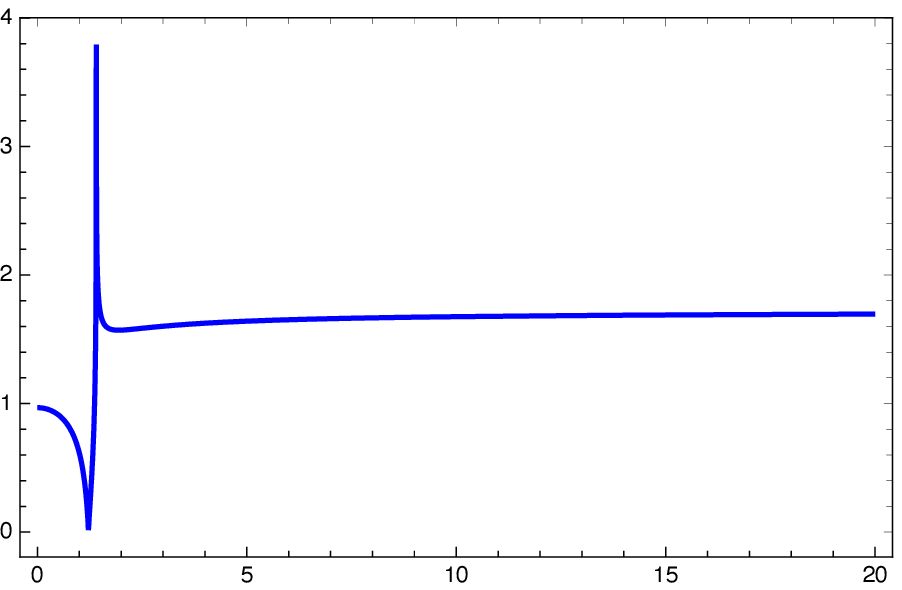}
	\put(-160,110){$\mathcal{D}$}
	\put(-90,-10){$\Delta/\omega_m$}
	\put(-90,110){\textbf{(a)}}
	~~~~~~~~\quad
	\includegraphics[scale=.64]{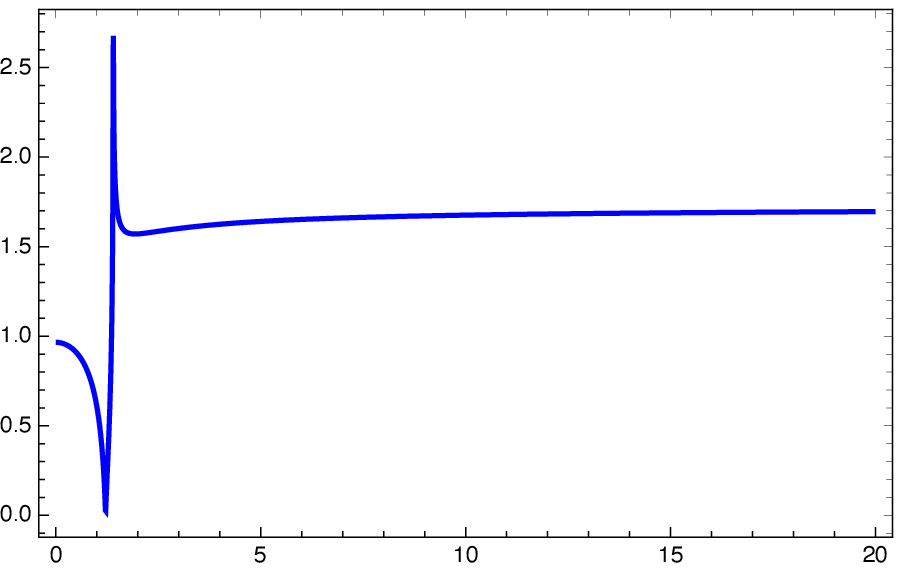}
	\put(-160,110){$\mathcal{D}$}
	\put(-90,-10){$\Delta/\omega_m$}
	\put(-90,110){\textbf{(b)}}\\
	\includegraphics[scale=.64]{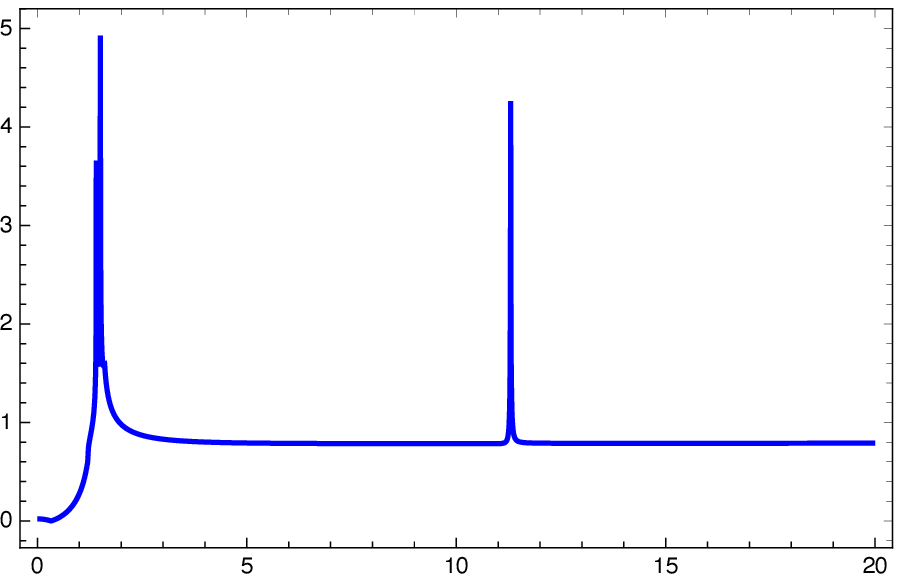}
	\put(-160,110){$\mathcal{D}$}
	\put(-90,-10){$\Delta/\omega_m$}
	\put(-90,110){\textbf{(c)}}
	~~~~~~~~\quad
	\includegraphics[scale=.64]{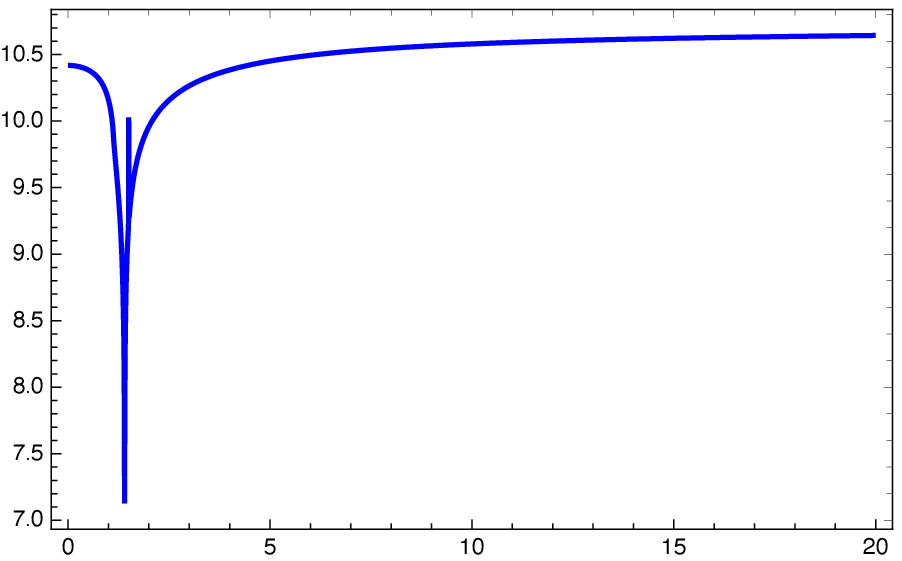}
	\put(-160,110){$\mathcal{D}$}
	\put(-90,-10){$\Delta/\omega_m$}
	\put(-90,110){\textbf{(d)}}
		\caption{Dynamics of Gaussian  quantum discord  versus the normalized detuning parameter $\Delta/\omega_m$, for BEC-second optical mode system.  We set,  $\kappa=14 \times 2 \pi $ MHz, $\mu^{A}=\mu^{B}=S_A=S_B=8 \times 2 \pi$ MHz, $\gamma_m=\gamma_{sm}=100 \times 2 \pi$ MHz, $\Omega=10 \times 2 \pi$. Moreover, (a)  $\bar{n}=N=0$, $J=1Hz$, (b) $\bar{n}=836, N=0$, $J=1Hz$, (c) $\bar{n}=N=0$, $J=0.5Hz$ and (d) $\bar{n}=14642, N=0.1$, $J=1Hz$.}
		\label{f7}
	\end{center}
\end{figure}

Finally, to acquire further insights into quantum correlation of the optomechanical system, we illustrate Fig.(\ref{f7} the dynamics of quantum correlation by means of Gauassian quantum discord when the BEC interacted with the optical resonator of the second optomechanical cavity. As is clear, we recover the same previous behavior as in Fig.(\ref{f6}) but with different amplitudes. Within this, we clearly see that the covariance matrix of this system, namely  $C_v^{8}$ is entangled ($\mathcal{D}>0$) for many critical values of the normalized detuning.\\

Overall, the peaks appearing in the behaviour of Gaussian quantum discord refer  to the transition pattern phenomenon \cite{Heymans}. Indeed, this transition  between quantum states occurs through the probabilistic pattern of energy emission or absorption. The transition pattern that emerges from these processes depends on several factors, including the energy difference between the initial and final states, the number and types of particles involved, and the geometry of the system. Importantly, the transition pattern can strongly influence the amount and dynamics of quantum discord. Indeed, the set of transition probabilities that describe the likelihood of the system to transit from one state to another, depends on the number and distribution of resonances in the system. Specifically, it is clear that our systems with higher levels of quantum discord tend to have more complex and nontrivial transition patterns, whereas low levels of quantum discord tend to have more simple and regular transition patterns. Hence, we conclude that the number of resonance transition patterns can affect the quantum discord between subsystems, particularly in systems with a large number of energy levels due to the passage of the BEC system inside the coupled hybrid optomechanical cavities. Close interpretation but with different systems was reported in Refs.\cite{TP1,TP2}. Therefore, from the obtained above results we can summarize the necessary conditions used to estimate the quality of correlations between BEC and the coupled hybrid optomechanical cavities. Indeed, we can conclude that one can control various optomechanical parameters. In general,  the corresponding covariance sub-matrices are correlated in particular when the Gaussian quantum discord is plotted with respect to changes in the normalized detuning parameter by keeping various initial settings of mean number of photons, strength coupling between cavities parameters.
	
	\section{Conclusion}
	
We presented a simple analysis of the dynamics of an interacted pair of a coupled optomechanical cavities, which we propose to be two Fabry-Pérot cavities of length $L$ with a moving end mirror. Moreover, we have supposed that both cavities are pumped by two-mode squeezed light sources. After exactly solving the set of  Langevin equations of the joint optomechanical systems, where  each hybrid cavity is interacted with a BEC, we have investigated the quantum correlation. Indeed, we have examined the Gaussian quantum discord between various subsystems of the whole ($12\times12$) covariance matrix, namely the first intracavity photon-phonon, second intracavity photon-phonon,  cavity modes, mechanical resonator modes, BEC-first mechanical mode, BEC-second mechanical mode, BEC-first optical mode and BEC-second  mechanical mode subsystems. We have concluded that all the Gaussian quantum discord quantifiers of the first and  second intracavity photon-phonon oscillate similarly between their upper and lower bounds since we have supposed that both optomechanical cavities are identical. We have found that the obtained covariance matrices are entangled for particular initial settings and for some specefic intervals of the normalized detuning parameter. By making a comparative study between the Gaussian quantum discord of the covariance matrices representing  the interaction between the pair hybrid optomechanical cavities and BEC, we have obtained some interesting outcomes. In general, we have gained an entangled covariance matrix related to the cavity modes, BEC-first mechanical mode, BEC-second mechanical mode, BEC-first optical mode and BEC-second  mechanical mode subsystems while the Gaussian quantum discord for the mechanical modes, first  and second intracavity photon-phonon subsystems can be entangled or unentangled since this measure of quantum correlation is lower than one. Indeed, we have concluded that the generation of quantum correlation and its  robustness depend basically on the physical parameters of the optomechanical system.  \\
	
	In summary, it is possible to prepare an entangled covariance matrix from the interaction between an interacted two-optomechanical system and BEC, with N atoms trapped in an optical lattice potential. These results can be utilized analytically and experimentally in many areas of quantum information theory and secure communication. Our future perspective will concern other optomechanical systems in order to extend the dimension of the  present work. Moreover, it will be motivated for us to investigate other kinds of quantum correlation measures to use them as a good resource in the quantum teleportation protocol and many other applications in quantum information theory. 

	\section*{Appendix A}
	
	Here, we give an exhaustive description  necessary to compute the whole  $(12\times12)$ covariance matrix as indicated at the end of section $3$.  To do this, let first solve the coupled Langevin equations of motion (\ref{Lang}). In fact, let's focus our attention on the linearization approach to solve these equations \cite{k8}. It consists of assuming that the operators can amended by small fluctuation from their steady state solutions, namely $\hat{a}=\alpha_s+\delta\hat{a}, \hat{b}=\beta_s+\delta\hat{b}, \hat{p}_{A(B)}=p_{as(bs)}+\delta\hat{p}_{A(B)}, \hat{q}_{A(B)}=q_{as(bs)}+\delta\hat{q}_{A(B)}$. Based on this and conserving only the linear terms, one can obtain the following coupled equations of motion for the fluctuation parts:
	\begin{eqnarray}\label{linearized}
		\frac{d\hat{\delta q}_A}{dt}&=&\omega_m^{A} \hat{\delta p}_A, \nonumber  \\
		\frac{d\hat{\delta q}_B}{dt}&=&\omega_m^{B} \hat{\delta p}_B, \nonumber  \\
		\frac{d\hat{\delta p}_A}{dt}&=&-\omega_m^{A} \hat{\delta q}_A-\gamma_{m}^{A}\hat{\delta p}_A+\mu_A \hat{\delta X}_A+\hat{I}_A, \nonumber  \\
		\frac{d\hat{\delta p}_B}{dt}&=&-\omega_m^{B} \hat{\delta q}_B-\gamma_{m}^{B}\hat{\delta p}_B+\mu_B \hat{\delta X}_B+\hat{I}_B, \nonumber  \\
		\frac{d\hat{\delta X}_A}{dt}&=&-\kappa_A \hat{\delta X}_A+\Delta_A \hat{\delta Y}_A+\sqrt{2\kappa_A} \hat{\delta X}_{in}^{A}-i\frac{J}{\sqrt{2}}(\beta_s-\beta_s^{*})+J\hat{\delta Y}_B,  \nonumber  \\
		\frac{d\hat{\delta X}_B}{dt}&=&-\kappa_B \hat{\delta X}_B+\Delta_B \hat{\delta Y}_B+\sqrt{2\kappa_B} \hat{\delta X}_{in}^{B}-i\frac{J}{\sqrt{2}}(\alpha_s-\alpha_s^{*})+J\hat{\delta Y}_A,\nonumber  \\
		\frac{d\hat{\delta Y}_A}{dt}&=&-\kappa_A \hat{\delta Y}_A+\Delta_A \hat{\delta X}_A+\sqrt{2\kappa_A} \hat{\delta Y}_{in}^{A}+S_A\hat{\delta q}_A-J(\beta_s+\beta_s^{*})-J\hat{\delta X}_B,  \nonumber  \\
		\frac{d\hat{\delta Y}_B}{dt}&=&-\kappa_B \hat{\delta Y}_B+\Delta_B \hat{\delta X}_B+\sqrt{2\kappa_B} \hat{\delta Y}_{in}^{B}+S_B\hat{\delta q}_B-J(\alpha_s+\alpha_s^{*})-J\hat{\delta X}_A,  \nonumber  \\
		\frac{d\hat{\delta Q}_A}{dt}&=&-\Omega_{A} \hat{\delta P}_A-\gamma_{sm}^{A}\hat{\delta Q}_A+\hat{I}_{2m}^{A}, \nonumber  \\
		\frac{d\hat{\delta Q}_B}{dt}&=&-\Omega_{B} \hat{\delta P}_B-\gamma_{sm}^{B}\hat{\delta Q}_B+\hat{I}_{2m}^{B}, \nonumber  \\
		\frac{d\hat{\delta P}_A}{dt}&=&-\nu_{A}+\Omega_{A} \hat{\delta Q}_A+\mu_A \hat{\delta X}_A-\gamma_{sm}^{A}\hat{\delta P}_A+\hat{I}_{1m}^{A}, \nonumber  \\
		\frac{d\hat{\delta P}_B}{dt}&=&-\nu_{B}+\Omega_{B} \hat{\delta Q}_B+\mu_B \hat{\delta X}_B-\gamma_{sm}^{B}\hat{\delta P}_B+\hat{I}_{1m}^{B}.
	\end{eqnarray}
	Here $\mu_{A(B)}=\sqrt{2}J\alpha_s(\beta_s)$ denotes  the effective optomechanical coupling parameter, whereas, $\alpha_s$ and $\beta_s$ are  real, while  $\nu_{A(B)}=\xi_m^{A}|\alpha_s(\beta_s)|^2-(\frac{\Omega_{A(B)}^2}{\gamma_{sm}^{A(B)}}+\gamma_{sm}^{A(B)})$ and $S_{A(B)}=2\sqrt{2}\xi_{A(B)}\alpha_s(\beta_s)$. In addition, we took into account the following quadratures as well as  input quadrature noise operators: 
	\begin{eqnarray}
		\delta \hat{X}_{A(B)}&=&\frac{\delta \hat{ a}_{A(B)}+\delta \hat{ a}_{A(B)}^{\dagger}}{\sqrt{2}},\nonumber  \\
		\delta \hat{Y}_{A(B)}&=&\frac{\delta \hat{ a}_{A(B)}-\delta \hat{ a}_{A(B)}^{\dagger}}{i\sqrt{2}},\nonumber  \\
		\delta \hat{X}_{A(B)}^{in}&=&\frac{\delta \hat{ a}_{A(B)}^{in}+\delta \hat{ a}_{A(B)}^{\dagger^{in}}}{\sqrt{2}},\nonumber  \\
		\delta \hat{Y}_{A(B)}^{in}&=&\frac{\delta \hat{ a}_{A(B)}^{in}-\delta \hat{ a}_{A(B)}^{\dagger^{in}}}{i\sqrt{2}}. 
	\end{eqnarray}
	
	It is obvious that linearized Langevin equations appeared in Eq.(\ref{linearized}) can be reconstruct in the convenient form:
	
	\begin{equation}\label{R}
		\dot{R}(t)=DR(t)+\mathcal{O}(t), 
	\end{equation}
	
	where $\mathcal{O}(t)$ denotes the noise vector.  It is defined as 
	
	\begin{eqnarray}
		\mathcal{O}(t)&=& \big(0, \sqrt{2\gamma_m^{A}}\hat{b}_{in}(t), \sqrt{2\kappa_{A}}\hat{X}_{in}^{A}(t),\sqrt{2\kappa_{A}}\hat{Y}_{in}^{A}(t), 0, \sqrt{2\gamma_{B}}\hat{c}_{in}(t), \sqrt{2\kappa_{B}}\hat{X}_{in}^{B}(t), \sqrt{2\kappa_{B}}\hat{Y}_{in}^{B}(t) \big)^{T}. 
	\end{eqnarray}
	
	Moreover, the matrix $D$ in Eq.(\ref{R})  denotes  the drift matrix. Indeed, by keeping all the operators in  Eq.(\ref{linearized}) except the last four equations since the atom is considered as an auxiliary system (environment) used to connect
	both cavities. Hence, the drift matrix $D$ is expressed as the following matrix form
	
	\begin{eqnarray}\label{Drift}
		D&=&\begin{pmatrix}
			 D_1&D_2\\
		 D_3&D_4\\
		\end{pmatrix}.
	\end{eqnarray}
where 
  \begin{eqnarray}\label{}
 	D_1&=&\begin{pmatrix}
 		0 & \omega_m^{A} & 0 & 0 & 0 & 0\\
 			- \omega_m^{A}    &  -\gamma_m^{A} &  \mu_{A}  &   0& 0 & 0\\
 				0 & 0 &-\kappa_A &   \Delta_A & 0 & 0\\
 				S_A& 0 &  \Delta_A  & -\kappa_A &0&0\\
 				0 &0& 0 & 0 & 0 &  \omega_m^{B}\\
 				0 &0& 0 & 0 & 0& - \omega_m^{B} \\
	\end{pmatrix}, \quad \quad 	D2=\begin{pmatrix}
0 & 0 &   0& 0 & 0 & 0\\
0 & 0&   0& 0 & 0 & 0\\
	 0 & J&   0& 0 & 0 & 0\\
	-J&0&   0& 0 & 0 & 0\\
	 0 & 0&   0& 0 & 0 & 0\\
-\gamma_m^{B} & \mu_{B}  & 0&   0& 0 & 0 \\
\end{pmatrix}\nonumber\\
	D3&=&\begin{pmatrix}
	0     & 0 &0 &   J & 0 & 0\\
0& 0 &  -J  & 0&S_B&0\\
0& 0 &  0  & 0&0&0\\
0& 0 & \mu^{A} & 0&0&0\\
	0& 0 &  0  & 0 &0 &0 \\
	0& 0 &  0  & 0&0&0\\
\end{pmatrix}, \quad \quad 	D4=\begin{pmatrix}
 -\kappa_B  &\Delta_B&   0& 0 & 0 & 0\\
\Delta_B&-\kappa_B&   0& 0 & 0 & 0\\
0&0&    -\gamma_{sm}^{A} &   \Omega^{A}  & 0 & 0\\
0&0&    \Omega^{A}  &   -\gamma_{sm}^{A}  & 0 & 0\\
0 &0 &   0& 0 &  -\gamma_{sm}^{B} &   -\Omega^{B}\\
\mu^{B}&0&   0& 0 & \Omega^{B}  & -\gamma_{sm}^{B}
\end{pmatrix}
\end{eqnarray}
	The interaction between the optical and mechanical resonators is characterized by a bipartite state with continuous variables. Let us examine the steady covariance matrix, namely $C_v$ via solving the following Lyapunov equation
	\begin{equation}\label{LE}
		DC_v+C_vD^{T}=-F, 
	\end{equation}
	where $F$ is diffusion matrix, is obtained by the noise correlation functions and by using Eq.(\ref{t}). The diffusion matrix is obtained as 	\begin{eqnarray}\label{F}
			F&=&\begin{pmatrix}
				F_1&F_2\\
				F_3&F_4\\
			\end{pmatrix}.
		\end{eqnarray}
		where 
		\begin{eqnarray}\label{}
			F_1&=&\begin{pmatrix}
				0 & 0& 0 & 0 & 0 & 0\\
			0   &  \gamma_m(1+2\bar{n}) &  0  &   0& 0 & 0\\
				0 & 0 &\kappa(1+2N) &   0 & 0 & 0\\
				0& 0 & 0  & \kappa(1+2N) &0&0\\
				0 &0& 0 & 0 & 0 &  0\\
				0 &0& 0 & 0 & 0& \gamma_m(1+2\bar{n})\\
			\end{pmatrix}, \quad \quad 	F2=\begin{pmatrix}
				0 & 0 &   0& 0 & 0 & 0\\
				0 & 0&   0& 0 & 0 & 0\\
				2M\kappa  & 0&   0& 0 & 0 & 0\\
				0&-2M\kappa&   0& 0 & 0 & 0\\
				0 & 0&   0& 0 & 0 & 0\\
				0 & 0  & 0&   0& 0 & 0 \\
			\end{pmatrix}\nonumber\\
			F3&=&\begin{pmatrix}
				0     & 0 &2M\kappa &   0& 0 & 0\\
				0& 0 &  0  & -2M\kappa&0&0\\
				0& 0 &  0  & 0&0&0\\
				0& 0 & 0& 0&0&0\\
				0& 0 &  0  & 0 &0 &0 \\
				0& 0 &  0  & 0&0&0\\
			\end{pmatrix},     \nonumber\\
			F4&=&\begin{pmatrix}
			\kappa(1+2N)   &0&   0& 0 & 0 & 0\\
				0&\kappa(1+2N) &   0& 0 & 0 & 0\\
				0&0&    \gamma_{sm}(1+2\bar{n})&  0  & 0 & 0\\
				0&0&   0  &   \gamma_{sm}(1+2\bar{n}) & 0 & 0\\
				0 &0 &   0& 0 &\gamma_{sm}(1+2\bar{n})&  0\\
			0&0&   0& 0 & 0  & \gamma_{sm}(1+2\bar{n})
			\end{pmatrix}.
	\end{eqnarray}	
	Now, since the drift and diffusion matrices are well known, one can   find the corresponding $(12\times12)$ covariance matrix via solving  numerically the Lyapunov equation in Eq. (\ref{LE}).

	\section*{Appendix B: Stability criterion}

	In this appendix, we show that parameters already chosen during our numerical simulations of quantum discord dynamics guarantees stability of the system. Indeed, we provide the stability analysis by means the Ruths–Hurwitz criterion \cite{Miskeen, Hammerer}. In fact the mean subject of this criterion is to determine the characters of the solution of the  following characteristic equation associated to the drift matrix  in Eq.(\ref{Drift}) as: 
	
	\begin{equation}
		S_{11} \lambda^{11}+S_{10} \lambda^{10}+....+S_0=0. 
	\end{equation}
	Hence, the Ruth-Hurwitz criterion sets that the system is be stable if and oly if the characters $S_i$ ($i=0,...11$) are positives. In this regards, by imposing  the parameters closed to the topical experiments used in our numerical simulations of quantum discord dynamics, we plot in the following figures versus the normalized detuning parameter $\Delta/\omega_m$. As it is clear from the figure,  the system is completely stable since $S_i>0$ ($i=0,...11$). 

	\begin{figure}
		\begin{center}
			\includegraphics[scale=.37]{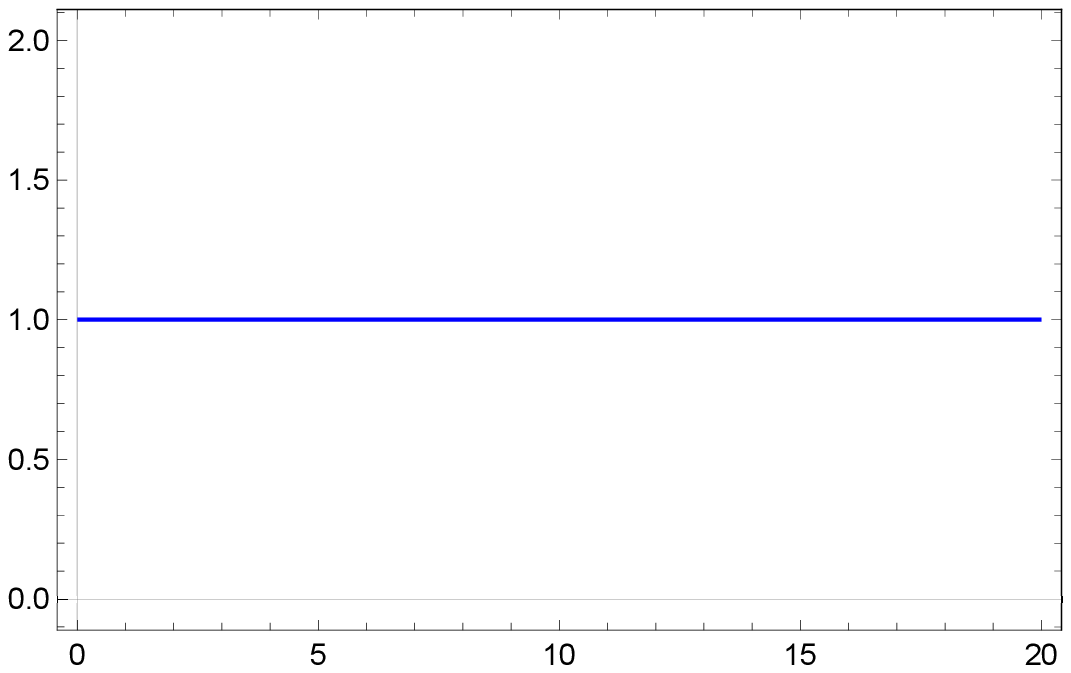}
			\put(-120,75){$S_0$}
		\put(-70,-13){$\Delta/\omega_m$}
			\put(-59,60){\textbf{(a)}}
			~~~~~~~~\quad
			\includegraphics[scale=.37]{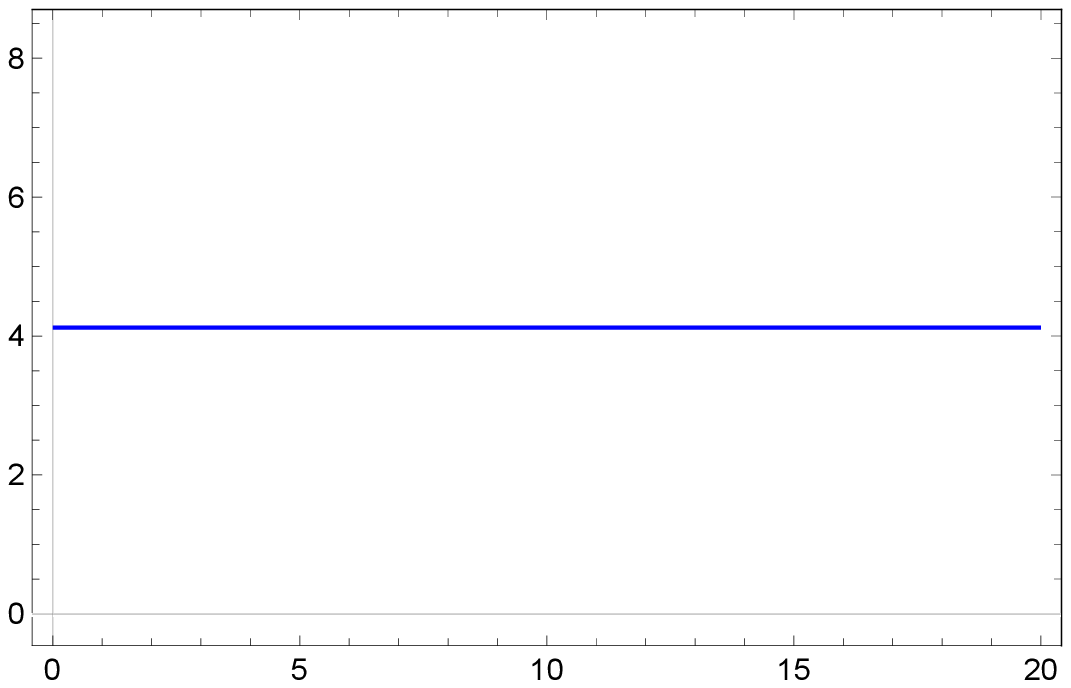}
			\put(-120,75){$S_1(\times 10^{9})$}
			\put(-70,-13){$\Delta/\omega_m$}
			\put(-59,60){\textbf{(b)}}
			~~~~~~~~\quad
			\includegraphics[scale=.37]{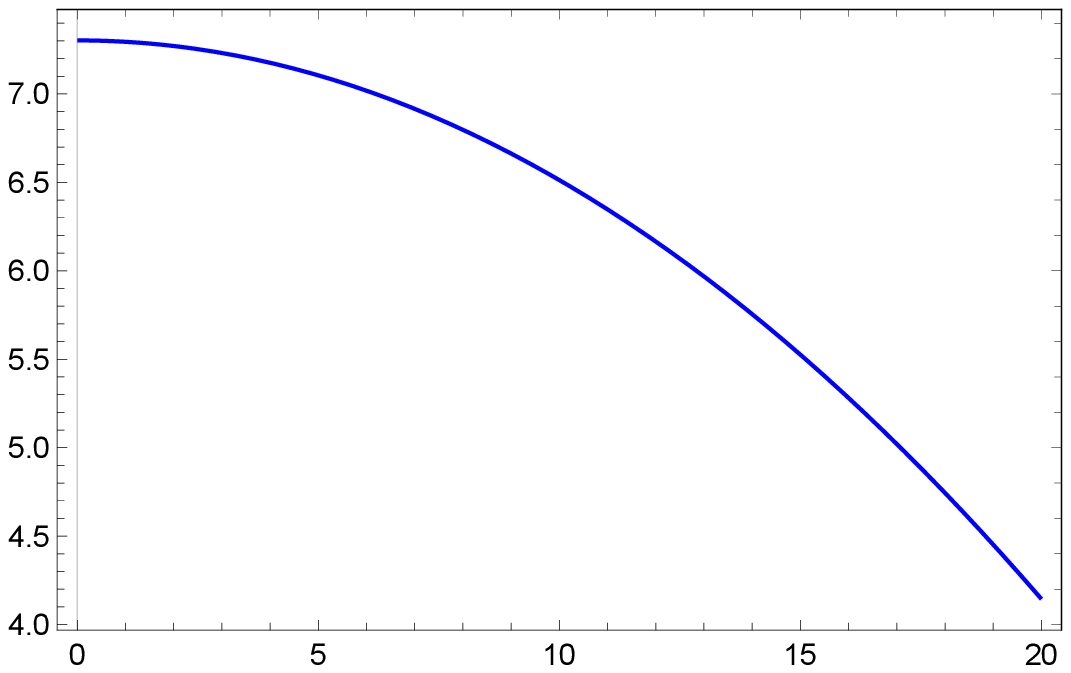}
			\put(-120,75){$S_2(\times 10^{18})$}
			\put(-70,-13){$\Delta/\omega_m$}
			\put(-59,60){\textbf{(c)}}
			~~~~~~~~\quad
			\includegraphics[scale=.37]{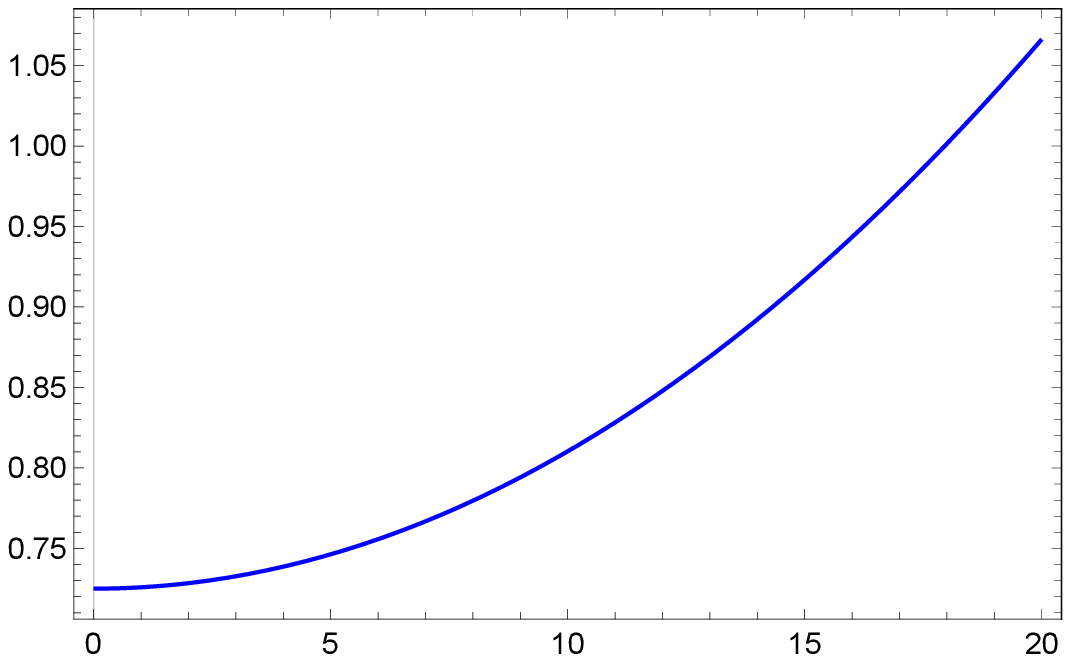}
			\put(-120,75){$S_3(\times 10^{28})$}
			\put(-70,-13){$\Delta/\omega_m$}
			\put(-59,60){\textbf{(d)}}\\
			\includegraphics[scale=.37]{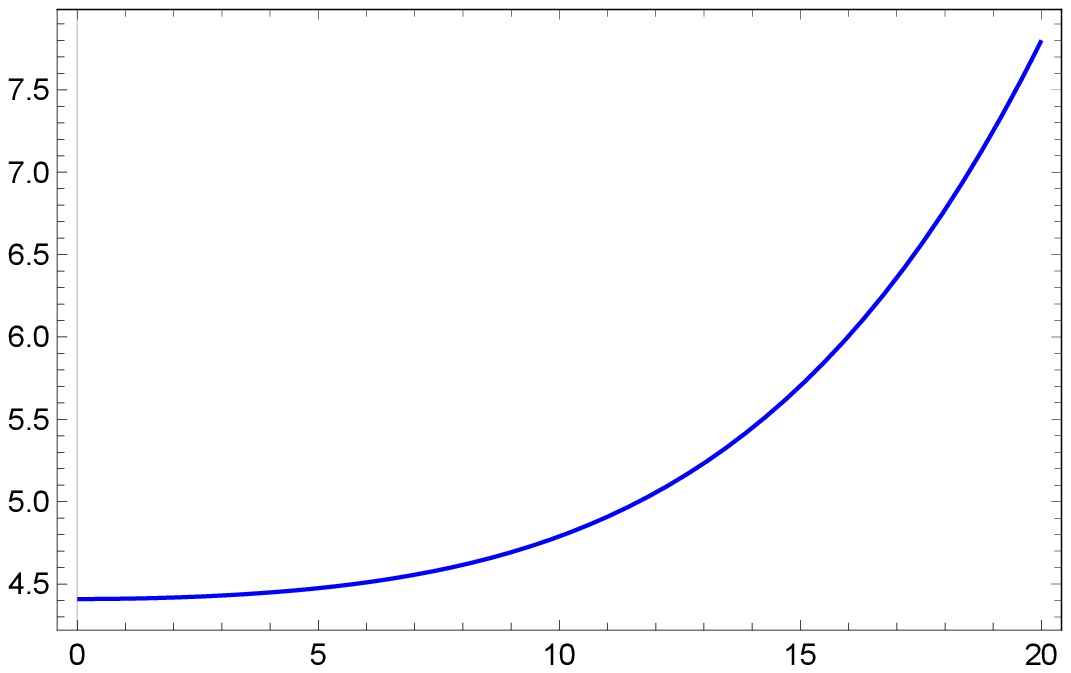}
			\put(-120,75){$S_4(\times 10^{36})$}
			\put(-70,-13){$\Delta/\omega_m$}
			\put(-59,60){\textbf{(e)}}
			~~~~~~~~\quad
			\includegraphics[scale=.37]{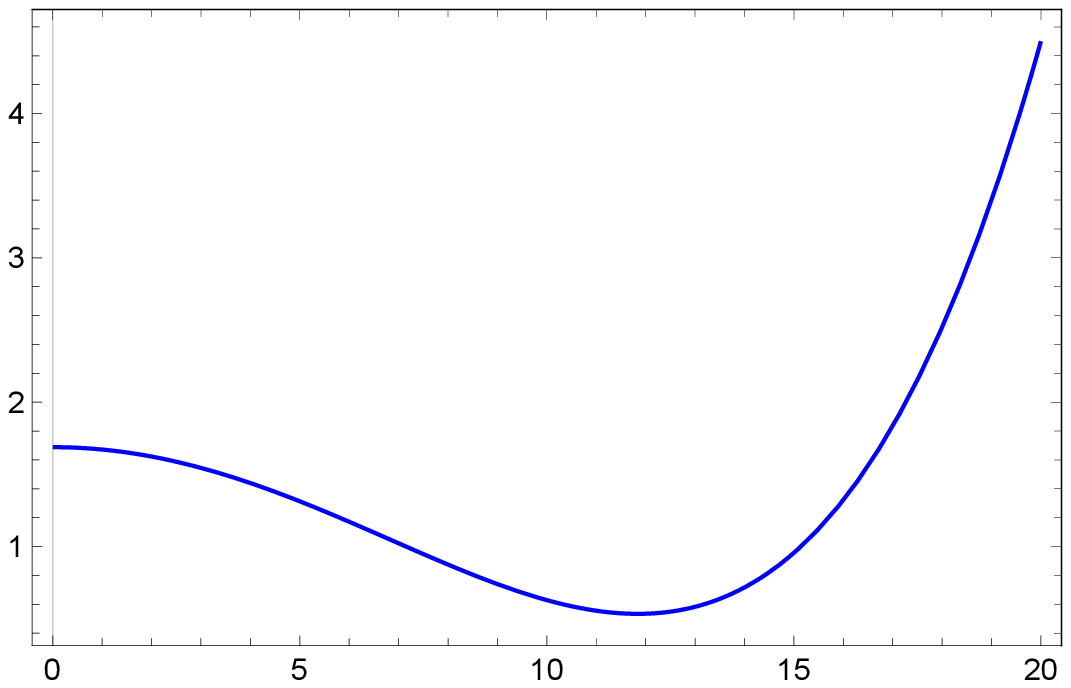}
			\put(-120,75){$S_5(\times 10^{45})$}
			\put(-70,-13){$\Delta/\omega_m$}
		\put(-59,60){\textbf{(f)}}
			~~~~~~~~\quad
			\includegraphics[scale=.37]{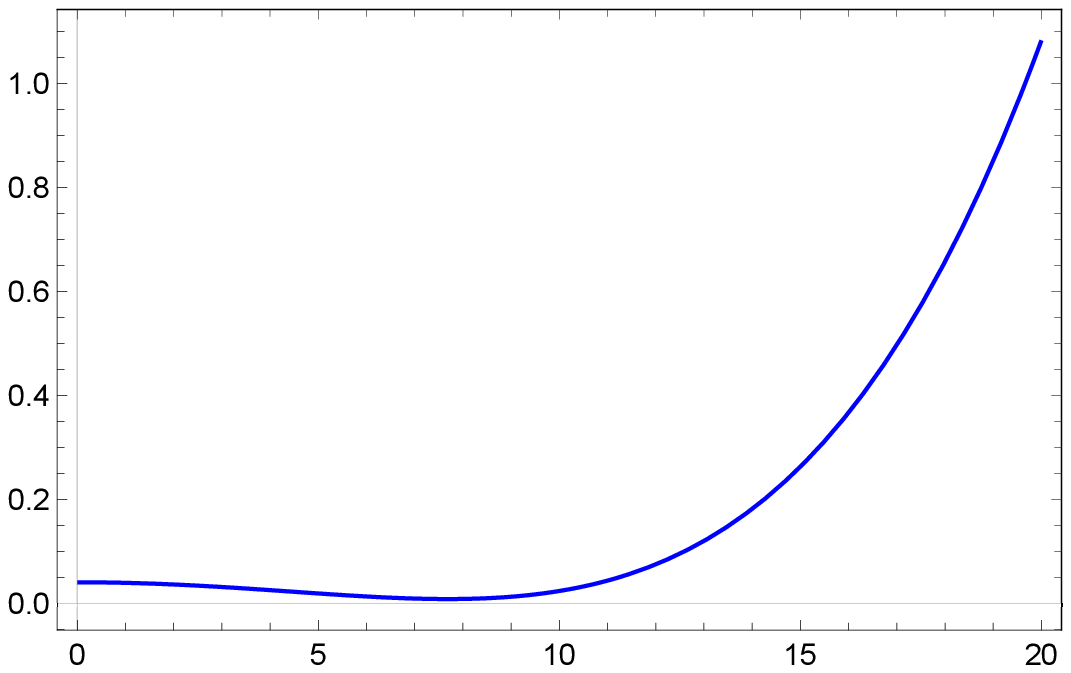}
			\put(-120,75){$S_6(\times 10^{55})$}
			\put(-70,-13){$\Delta/\omega_m$}
			\put(-59,60){\textbf{(g)}}
			~~~~~~~~\quad
			\includegraphics[scale=.37]{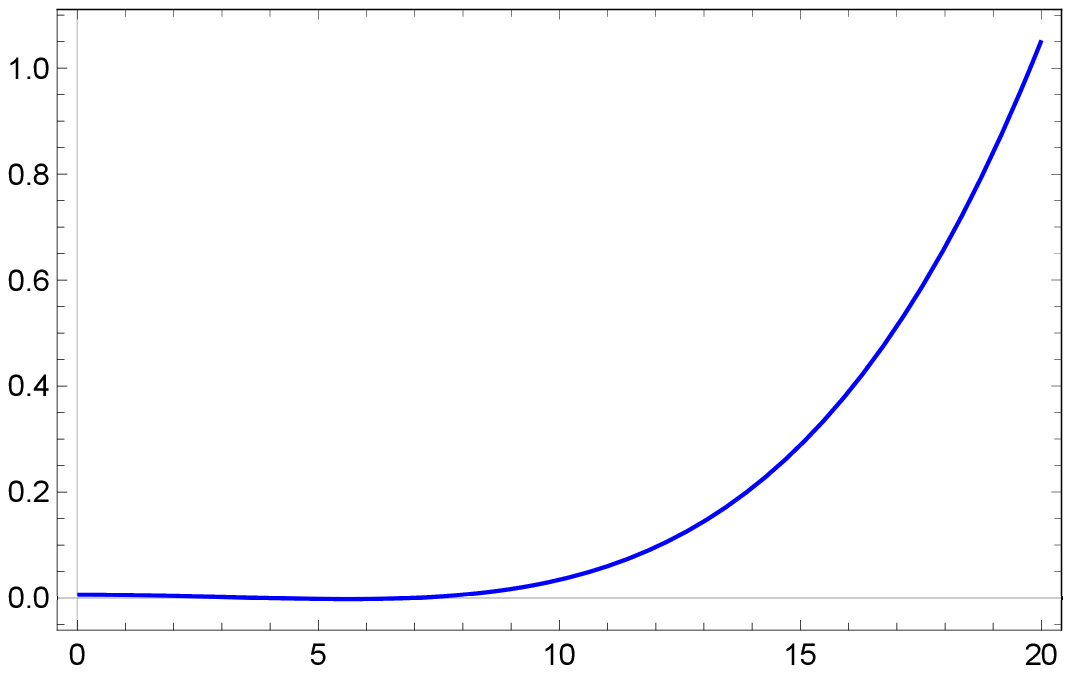}
			\put(-120,75){$S_7(\times 10^{64})$}
			\put(-70,-13){$\Delta/\omega_m$}
			\put(-59,60){\textbf{(h)}}\\
			\includegraphics[scale=.37]{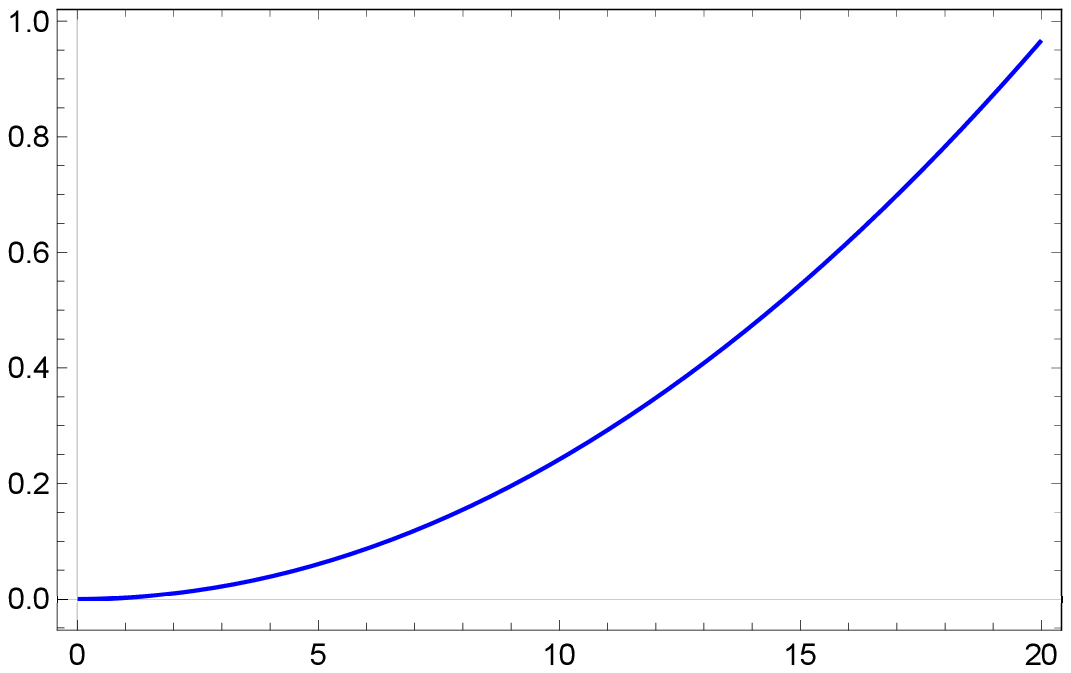}
			\put(-120,75){$S_8(\times 10^{127})$}
			\put(-70,-13){$\Delta/\omega_m$}
			\put(-59,60){\textbf{(i)}}
			~~~~~~~~\quad
			\includegraphics[scale=.37]{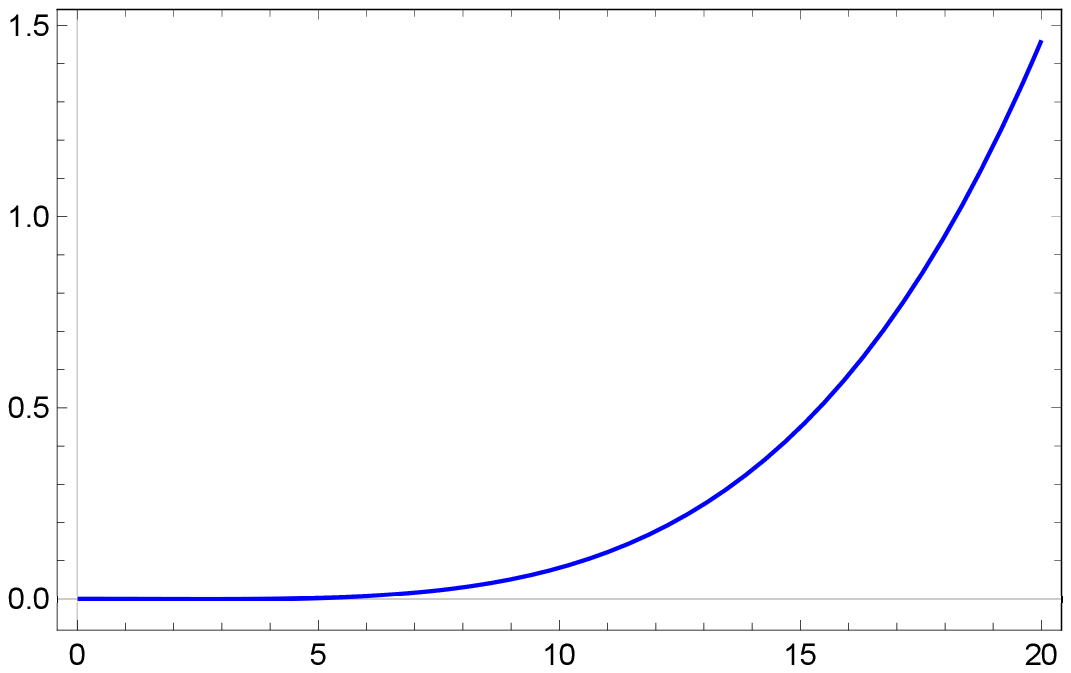}
			\put(-120,75){$S_9(\times 10^{81})$}
			\put(-70,-13){$\Delta/\omega_m$}
			\put(-59,60){\textbf{(j)}}
			~~~~~~~~\quad
			\includegraphics[scale=.37]{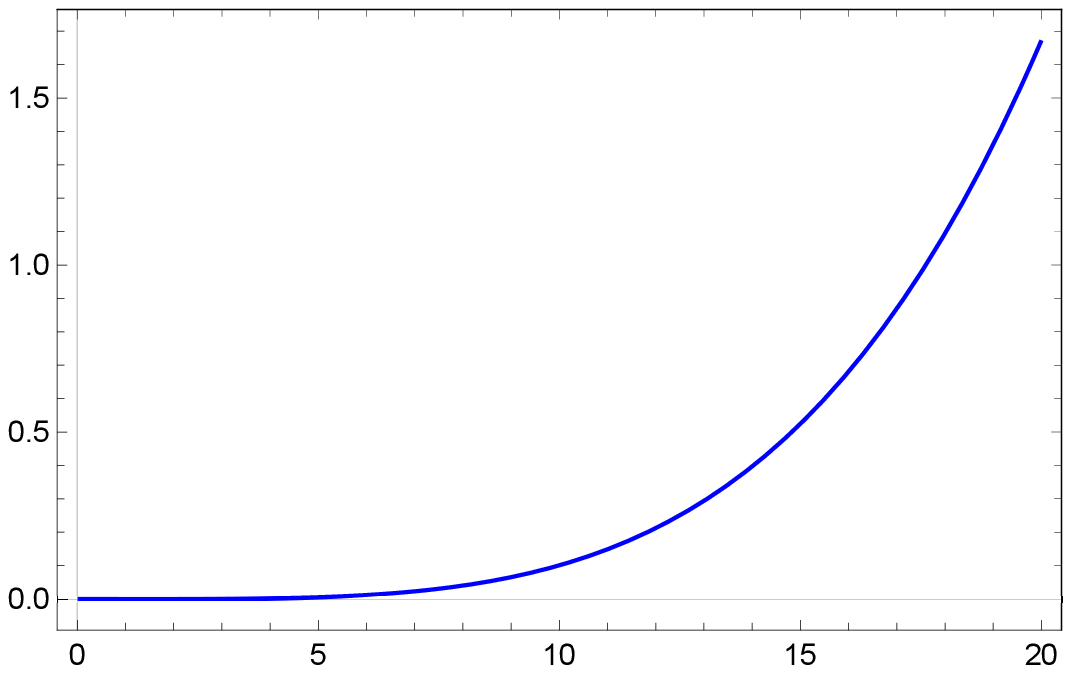}
			\put(-120,75){$S_{10}(\times 10^{89})$}
			\put(-70,-13){$\Delta/\omega_m$}
			\put(-59,60){\textbf{(k)}}
			~~~~~~~~\quad
			\includegraphics[scale=.37]{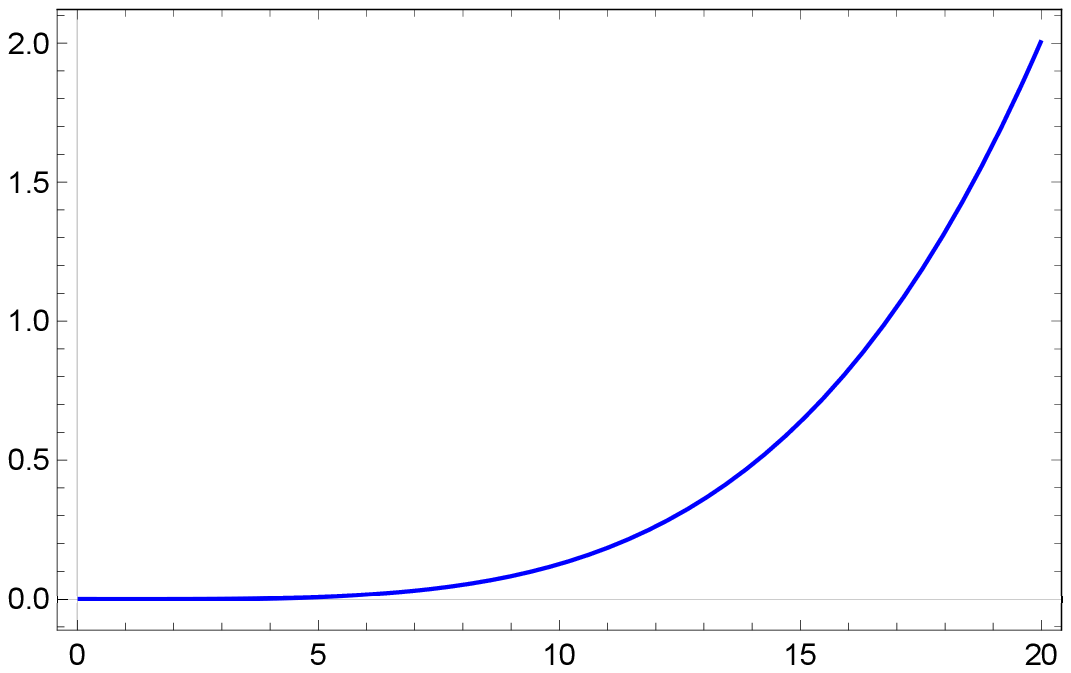}
			\put(-120,75){$S_{11}(\times 10^{96})$}
			\put(-70,-13){$\Delta/\omega_m$}
			\put(-59,60){\textbf{(l)}}
			\caption{The stability conditions versus the normalized parameter $\Delta/\omega_m$ where $\kappa=14 \times 2 \pi$ MHz, $\mu^{A}=\mu^{B}=S_A=S_B=8 \times 2 \pi$ MHz, $\gamma_m=\gamma_{sm}=100 \times 2 \pi$ MHz, $\Omega=10 \times 2 \pi$ and $J=1 Hz$.}
			\label{f1app}
		\end{center}
	\end{figure}
	
		\newpage

	\section*{Declaration of Interest}
	
	The authors declare that they have no conflict of interest. 
	
	\section*{Data availability statement}
	
	No data statement is available.

\end{document}